\documentclass[nobibnotes,nofootinbib,twocolumn]{revtex4}
\usepackage[english]{babel}
\usepackage{ucs}
\usepackage[utf8x]{inputenc}
\usepackage{amsmath,amssymb,bm,cancel}
\usepackage{graphicx}
\usepackage{subfigure}
\usepackage{color}
\graphicspath{{figures/}}
\def \grad {\boldsymbol{\nabla}}
\def \div {\grad\cdot}
\def \curl {\grad\times}
\newcommand{\code}[1]{\texttt{#1}}

\begin{document}
\title{Hybrid guiding-centre/full-orbit simulations in
  non-axisymmetric magnetic geometry exploiting general criterion for
  guiding-centre accuracy}
\author{D Pfefferlé$^1$, J P Graves$^1$, W A Cooper$^1$}
\affiliation{$^1$ École Polytechnique Fédérale de Lausanne (EPFL),
  Centre de Recherches en Physique des Plasmas (CRPP), CH-1015
  Lausanne, Switzerland}
\begin{abstract}
To identify under what conditions guiding-centre or full-orbit tracing
should be used, an estimation of the spatial variation of the magnetic
field is proposed, not only taking into account gradient and curvature
terms but also parallel currents and the local shearing of
field-lines. The criterion is derived for general three-dimensional
magnetic equilibria including stellarator plasmas. Details are
provided on how to implement it in cylindrical coordinates, and in
flux coordinates that rely on the geometric toroidal angle. A means of
switching between guiding-centre and full-orbit equations at first
order in Larmor radius with minimal discrepancy is shown. Techniques
are applied to a MAST (Mega Amp Spherical Tokamak) helical core
equilibrium in which the inner kinked flux-surfaces are tightly
compressed against the outer axisymmetric mantle and where the
parallel current peaks at the nearly rational surface. This is put in
relation with the simpler situation
$\bm{B}(x,y,z) = B_0 [\sin(kx) \bm{e_y} + \cos(kx)\bm{e_z}]$, for
which full orbits and lowest order drifts are obtained
analytically. In the kinked equilibrium, the full orbits of NBI fast
ions are solved numerically and shown to follow helical drift
surfaces. This result partially explains the off-axis redistribution
of NBI fast particles in the presence of MAST Long-Lived Modes (LLM).


\end{abstract}
\maketitle
\section{Introduction}
\label{sec:intro}
Fast ions play an important role in tokamaks and other fusion
devices. Ohmic heating being insufficient to reach ignition because
the collisionality in the plasma decreases with its temperature,
additional heating systems are envisaged. Neutral beam injection (NBI)
or ion cyclotron resonance heating (ICRH) will create significant
populations of energetic particles which, via collisional processes
and over slowing-down timescales, will transfer energy to the bulk
plasma. The efficient design of those heating systems requires the
accurate modelling of the dynamics of fast ions. At the level of
hybrid kinetic-MHD, the interaction between supra-thermal populations
and the background plasma depends on the individual motion of
energetic particles inside the magnetic geometry. Analytic models are
useful to understand the physics, however to make contact with more
realistic phenomena, numerical approaches are better suited. Resolving
the full-orbit of particles in strong magnetic fields is rather
strenuous, sometimes superfluous. Most approaches therefore solve the
guiding-centre drift equations (GCDE). The latter can be derived from
full-orbit equations by truncating their Taylor expansion in Larmor
radius \cite{northrop}. At any given order, Hamiltonian properties can
be recovered by retaining selected higher-order terms
\cite{boozer-1980}. A systematic derivation of GCDE is elegantly
obtained via Lie perturbation theory of the non-canonical phase-space
Lagrangian \cite{littlejohn-1983}. In most orbit simulations,
first-order GCDE are used on the basis that the magnetic field is
slowly varying with respect to the Larmor radius. There are
situations, especially in three-dimensional configurations, where the
field variation is large and where the primary assumptions of the
guiding-centre approximation are not cleanly satisfied. Being able to
quantify the field variation is important in order to assess the
applicability of GCDE.

In this paper, a numerical criterion is proposed for anticipating
where the magnetic field has large variations with respect to Larmor
radius, taking into account not only gradients and curvature but the
total perpendicular variation. The resulting criterion is applied to
NBI fast particles in 3D helical core equilibria. In these magnetic
configurations, the inner flux-surfaces are firmly kinked against the
outer axisymmetric mantle. Strong parallel currents develop at the
radial position where the q-profile is closest to unity. The field
variation in the helical core is significant enough to challenge first
order GCDE so that full-orbits must be studied to confirm the
transport properties that were established in previous work
\cite{pfefferle-nf}. In the latter, the redistribution of NBI fast
ions in the area where the flux surfaces are uncompressed was
explained by 3D guiding-centre drift motion. It must be verified that
the drift effects caused by the helical geometry are still true for
the full particle motion.

The paper is organised as follows. In
section~\ref{sec:field_variation}, the field variation criterion for
inspecting the limits of first-order GCDE is explained. Details are
given on how this criterion can be measured for flux coordinates with
the geometrical toroidal angle as well as for cylindrical
coordinates. In section~\ref{sec:switching}, full-orbit equations in
general curvilinear coordinates are recalled. A procedure to switch
between particle and guiding-centre variables with minimal discrepancy
between the two phase-spaces is shown. In
section~\ref{sec:heli_analysis}, the variation of the magnetic field
in an extreme case of MAST helical core is evaluated. From the
full-orbits of a given set of fast particles, the position of the
guiding-centres is verified to depict the same drift surfaces as with
GCDE. Focus is brought on deeply passing particles for which
guiding-centre and full-orbit trajectories are in principle the same.


%
\section{Estimation of magnetic field variations}
\label{sec:field_variation}
The guiding-centre approximation is often loosely justified by
assuming that the scale length of the magnetic field is large compared
with the gyroradius, or simply that gradients in the magnetic strength
are weak, $B/|\grad_\perp B|\gg \rho_L$. It is reminded that there are
situations where gradients of $B$ and even curvature are absent, but
the variation of the field can still be strong. For example, the
following magnetic field in Cartesian coordinates\footnote{It is
  immediately verified that this is a valid magnetic field,
  i.e. $\div\bm{B} = 0$.}
\begin{equation}
\label{eqn:sheared_magfield}
  \bm{B}(x,y,z) = B_0 [\sin(kx) \bm{e_y} + \cos(kx)\bm{e_z}].
\end{equation}
is such that its modulus, $B$, is constant and
$\curl\bm{B}=k\bm{B}$. Hence, there are no gradients, $\grad B =
\bm{0}$, and no curvature (see \cite{dHaeseleer} for definition)
\begin{align*}
  \bm{\kappa}&=(\bm{b}\cdot\grad)\bm{b}=-\bm{b}\times(\curl\bm{b}) \\
&= -\frac{\bm{B}}{B^2}\times(\curl\bm{B})+\frac{\bm{B}}{B^3}\times(\grad B\times\bm{B}) = \bm{0}
\end{align*}
where $\bm{b} = \bm{B}/B$. If the direction of the magnetic field is
rapidly changing with respect to the coordinate $x$ and if this
variation compares with the particle's Larmor radius, the first order
GCDE break down. Indeed, for this particular magnetic field, the
guiding-centre motion is described by $\dot{v}_{||} = 0$ and
$\dot{\bm{X}}= v_{||} \bm{b}$ and does not depend on the value of the
shearing parameter $k$. The full particle motion, which can be solved
analytically (see appendix~\ref{sec:orbit_sheared_field}), is strongly
dependent on $k$ in the sense that above a certain threshold (see
equation (\ref{eqn:shear_threshold})), the motion is not even along
the initial parallel axis. If $k$ is small, it is found that the
average drift velocity for a particle of mass $m$ and charge $q$
actually approaches
\begin{equation*}
  \bm{b}\cdot<\dot{\bm{x}}> = \bm{b}\cdot\dot{\bm{X}} \overset{k\rightarrow 0}{\longrightarrow}
  v_{||} + \frac{kv_\perp^2}{4\omega_0}
\end{equation*}
where $v_{||}$ is the initial parallel velocity of the particle,
$v_\perp$ its initial perpendicular velocity and $\omega_0 = q B_0/m$
the gyro-frequency. This shows that the guiding-centre does not simply
follow $v_{||}$, but undergoes a drift even though $\grad B = 0$ and
$\bm{\kappa}=0$, due to the shearing of field-lines. This effect is
consistent with the so-called Baños drift \cite{banos} appearing at
second order in the guiding-centre expansion (see equation
(\ref{eqn:2nd_lagrangian}) and associated text).

GCDE are typically derived by expanding the magnetic field around the
position of the guiding-centre $\bm{x}=\bm{X}+\bm{\rho_\perp}$, where
$\bm{x}$ is the position of the particle, $\bm{X}$ the position of the
guiding-centre and $\bm{\rho_\perp}$ a vector of the size of the
Larmor radius perpendicular to the magnetic field. At first order, the
magnetic field is approximated as
\begin{equation*}
  \bm{B}(\bm{x}) \approx \bm{B}(\bm{X}) + (\bm{\rho_\perp}\cdot\grad)\bm{B}(\bm{X}).
\end{equation*}
Such an expansion can be qualified as reasonable as long as the
variation of the field is much smaller than the field itself, i.e.
\begin{equation*}
  \frac{\left|\delta\bm{B}\right|}{|\bm{B}|} = 
  \frac{\left|\bm{B}(\bm{x})-\bm{B}(\bm{X})\right|}{|\bm{B}|} \approx
  \frac{\left|(\bm{\rho_\perp}\cdot\grad)\bm{B}\right|}{|\bm{B}|} \sim O(\epsilon) \ll 1.
\end{equation*}
This criterion should be respected when using first-order
GCDE. Hereafter, an explicit computation of the field variation is
detailed for general magnetic fields and in curvilinear coordinates.

First, it is useful to express the Larmor radius as the perpendicular
projection of a random vector, $\bm{\rho_\perp} = P \bm{\rho} =
(I-\bm{b}\bm{b})\bm{\rho}$. Then, the idea is to view the desired
scalar quantity as a bi-linear form
\begin{align*}
K(\bm{\rho},\bm{X})&=\left|(\bm{\rho_\perp}\cdot\grad)\bm{B}\right|^2 \\
  &= [(P\bm{\rho})\cdot\grad\bm{B}]\cdot[\grad\bm{B}\cdot (P\bm{\rho})] 
  = \hat{\rho}^i \hat{M}_{ij} \hat{\rho}^j
\end{align*}
where Einstein summation rule applies (and will apply by default
throughout the paper). The \emph{hat} notation is to stress that the
components $\hat{\rho}_i$ are those of the vector $\bm{\rho}$
expressed in an orthonormal (Cartesian) basis in the lab frame. The
importance of this statement will become clear later. The spectral
theorem states that an orthonormal basis of eigenvectors always exists
for $M$. In other words, $\bm{\rho}$ becomes an eigenvector if rotated
correctly. This also means that the maximum of
$K(\bm{\rho},\bm{X})$ with respect to $\bm{\rho}$, i.e. the
maximum eigenvalue $\lambda_{max}$ of matrix $M$, will correspond to
the maximum variation of the magnetic field in the perpendicular
direction. In standard Cartesian coordinates $(x^1,x^2,x^3)=(x,y,z)$,
$M$ is written
\begin{equation*}
  \hat{M}_{ij} =
  \hat{P}^m_i\frac{\partial\hat{B}_l}{\partial x^m} \frac{\partial\hat{B}^l}{\partial x^k} \hat{P}^k_j =
  [P^T D^T D P]_{ij}
\end{equation*}
where $\hat{B}^i=\hat{B}_i$ are the Cartesian components of the
magnetic vector field and
$\hat{P}^i_j = \delta^i_j - \frac{\hat{B}^i\hat{B}_j}{B^2}$. For the
purely-sheared magnetic field example given in equation
(\ref{eqn:sheared_magfield}), the Jacobian matrix $\hat{M}$ reduces to
\begin{equation*}
  \hat{M}_{ij} = k^2 B_0^2
  \begin{pmatrix}
    1 & 0 & 0\\
    0 & 0 & 0 \\
    0 & 0 & 0
  \end{pmatrix},
\end{equation*}
and the maximum eigenvalue is $\lambda_{max} = k^2B_0^2$. This implies
that a linear expansion in Larmor radius does notice the shearing of
field-lines and the presence of parallel currents.

Representing the field and its derivatives in Cartesian coordinates is
not always convenient because those coordinates are not native to the
geometry of the system. Cylindrical or flux coordinates are preferred
for applying this criterion in the case of toroidal magnetic fields
that are relevant to tokamak and stellarator physics. If
$(u^1,u^2,u^3)$ are those coordinates, $\grad\bm{B}$ is a covariant
derivative
\begin{align*}
  \grad\bm{B} &=\frac{\partial}{\partial u^j} (B^i\bm{e}_i) \grad u^j=
  \frac{\partial B^i}{\partial u^j}\bm{e}_i \grad u^j + B^k \frac{\partial \bm{e_k}}{\partial u^j}\grad u^j \\
  &= (\partial_j B^i + \Gamma_{jk}^iB^k)\bm{e}_i \grad u^j=B^i_{;j} \bm{e}_i\grad u^j
\end{align*}
or alternatively
\begin{align*}
  \grad\bm{B} & =B_{i;j}\grad u^i\grad u^j = (\partial_j B_i - \Gamma^k_{ij}B_k)\grad u^i \grad u^j\\
  & =(\partial_j B_i - \Gamma_{ij,k}B^k)\grad u^i \grad u^j
\end{align*}
where $\bm{e}_i=\frac{\partial \bm{x}}{\partial u^i}$ the covariant
basis, $\grad u^i$ the contravariant basis, $B^i$ the contravariant
and $B_i$ covariant components of the magnetic field, $\Gamma_{ij,k}$
is recognised as the Christoffel symbol of first type and
$\Gamma^i_{jk} = g^{il}\Gamma_{jk,l}$ of second type. The scalar
product in curvilinear coordinates is now expressed as a contraction
with the metric tensor, $\bm{a}\cdot\bm{b} = a^ib^j g_{ij} = a_i b_j
g^{ij} = a_i b^i = a^ib_i$. The tensor M becomes (multiple
possibilities)
\begin{equation*}
  M^{ij} = P^{ik} B^l_{;k} g_{lm} B^m_{;n} P^{nj} = P^{ik} B_{l;k} g^{lm} B_{m;n} P^{nj}
\end{equation*}
with $P^{ab} = g^{ab} - \frac{B^aB^b}{B^2}$.

If the $M$ tensor is diagonalised in the curvilinear basis, it will
yield the amplitude of the variations along a non-orthonormal basis,
which is not desired. The tensor must first be transformed to an
orthonormal basis before diagonalisation
\begin{equation*}
  K(\bm{\rho},\bm{X}) = 
  \hat{\rho}^k \bm{\hat{e}}_k \cdot \bm{e}_i M^{ij} \bm{e}_j \cdot \bm{\hat{e}}_l \hat{\rho}^l  =
  \hat{\rho}^k \frac{\partial x_k}{\partial u^i} M^{ij} \frac{\partial x_l}{\partial u^j} \hat{\rho}^l.
\end{equation*}
The relevant matrix to diagonalise is thus composed of the following
product of matrices
\begin{equation}
  \hat{M}_{ij} = [\Lambda P^T D^T G D P \Lambda^T]_{ij} = [V^T G V ]_{ij}
\end{equation}
with $V = DP\Lambda^T$, $\Lambda_{ij} = \frac{\partial x_i}{\partial
  u^j}$, $D^i_{\ j}=B^i_{;j}$ and $G_{ij}=g_{ij}$ (or $D_{ij}=B_{i;j}$
and $G^{ij}=g^{ij}$). The maximum eigenvalue of $\hat{M}$,
$\lambda_{max}(\bm{X}) = \text{max}[\text{eig}(\hat{M})]$, will then
provide the local estimate of the maximum variation of the magnetic
field in Cartesian coordinates. It is straightforward to find the
eigenvalues of $\hat{M}$. One is zero due to the projection in the
perpendicular direction, i.e.  $det(\hat{M})=0$. The remaining pair
are obtained via the quadratic equation
$\lambda^2-\textrm{trace}(\hat{M})\lambda + b = 0$,
\begin{equation*}
  \lambda_\pm = \frac{1}{2}\textrm{trace}(\hat{M}) \pm \frac{1}{2}\sqrt{\textrm{trace}(\hat{M})^2 - 4 b}
\end{equation*}
where $b = M_{11}M_{22} - M_{12}^2 + M_{11}M_{33} - M_{13}^2 +
M_{22}M_{33} - M_{23}^2$. $\lambda_+$ is evidently the maximum
eigenvalue and is noticed to be bound by $\textrm{trace}(\hat{M})$
(quicker to evaluate).

Finally, putting all the pieces together, the criterion that must
verify first-order GCDE is
\begin{multline}
\label{eqn:gc_condition}
\frac{\left|(\bm{\rho_\perp}\cdot\grad)\bm{B}\right|}{B} = 
\frac{\sqrt{\lambda_{max}}\rho_\perp}{B} =
\frac{m}{q}\frac{v_\perp}{B^2}\sqrt{\lambda_{max}} =\\
\sqrt{\frac{2m}{q}\frac{\mathcal{H}_\perp}{q}}\frac{\sqrt{\lambda_{max}}}{B^2}
= \sqrt{\frac{2\lambda_{max}}{B^3} \frac{m}{q} \frac{\mu}{q}} \ll 1.
\end{multline}
where $m$ is the particle mass, $q$ its charge,
$v_\perp =\rho_\perp qB/m$ the perpendicular velocity,
$\mathcal{H}_\perp = \frac{1}{2}m v_\perp^2$ the energy in the
perpendicular direction and
$\mu = m v_\perp^2/2 B = \mathcal{H}_\perp/B$ the magnetic moment,
which is an adiabatic constant of the particle motion.

Applied to toroidal devices where the magnetic field strength
$B\approx B_0R_0/R$, the field variation approximately corresponds to
the gradient of the magnetic field, which is
$\sqrt{\lambda_{max}} \approx B_0/R_0$ on axis. Considering fusion
alphas (${}^4$He$^{+2}$) for example, the factor
$\sqrt{2m/q}\approx 2.04\cdot 10^{-4}$ [kg/C]$^{1/2}$ and
$\sqrt{\mathcal{H}_\perp/q} = \sqrt{3.5\text{ MeV}/2} = 1.323\cdot
10^{-3}$
[eV]$^{1/2}$ so criterion (\ref{eqn:gc_condition}) coarsely
corresponds to $0.27/B_0R_0 \ll 1 $. The ratio $0.27/B_0R_0$
represents a few percent in tokamaks with a large major radius and a
strong magnetic field like ITER. In spherical tokamaks like MAST where
$R_0=0.8$m and $B_0=0.5$T, the variation length-scale due to gradients
corresponds to $10\%$ of the Larmor radius for alpha particles at
$80$keV. As it is argued in this paper, not only gradients give rise
to field variations but also curvature, parallel currents and the
shearing of field-lines. Therefore, the value of
$\sqrt{\lambda_{max}}$ is higher than $|\grad B|$ and the range of
``valid'' energies for first-order guiding-centre theory is all the
more reduced (i.e. quadratically).

Applied to the purely-sheared magnetic field of equation
(\ref{eqn:sheared_magfield}), the criterion reads
\begin{equation*}
k\rho_\perp = \frac{kv_\perp}{\Omega_C}\ll 1,
\end{equation*}
meaning that the error between the guiding-centre and the average
particle trajectory is of order $kv_\perp/\Omega_C$, as expected from
the discussion of full-orbits in appendix
\ref{sec:orbit_sheared_field} and the drift ordering.

A list of convenient expressions of covariant derivatives and
projectors is derived in curvilinear coordinates in the appendix
\ref{sec:fieldvariation_cyl} for the application to general plasma
equilibrium fields. In section \ref{sec:switching}, criterion
(\ref{eqn:gc_condition}) is used as a trigger to switch between GCDE
and full-orbit equations. It is also applied in section
\ref{sec:heli_fieldvariation} to the case of NBI fast particles in a
MAST helical core equilibrium in order to inspect the regions of
strong field variation and reveal the features of this particular
magnetic configuration.

%
\section{Guiding-centre to full-orbit equations switching}
\label{sec:switching}
The derivation of first-order guiding-centre equations via
non-canonical phase-space Lagrangian principles will not be repeated,
as the reader may prefer referring to
\cite{littlejohn-1983,cary-brizard}, applied to general straight
field-line coordinates to \cite{white-chance}, and to curvilinear
coordinates to \cite{pfefferle-nf}. The less common derivation of
full-orbit equations in curvilinear coordinates is briefly
recalled. The starting point is the standard Lagrangian of a charged
particle in a general electromagnetic field
\begin{align}
\mathcal{L}(u^i,v^i,t) &= \frac{1}{2}m \bm{v}^2 + q\bm{A}(u^i,t)\cdot\bm{v} - q\Phi_E(u^i,t)\nonumber\\
& = \frac{1}{2}m g_{mn} v^m v^n + q A_k v^k - q \Phi_E
\end{align}
where $(u^1,u^2,u^3)$ are the particle's coordinates,
$v^i = \dot{u}^i= \dot{\bm{x}}\cdot \grad u^i$ the contravariant
components of its velocity, $m$ the particle mass, $q$ the particle
charge,
$g_{mn}=\frac{\partial \bm{x}}{\partial u^m}\cdot\frac{\partial
  \bm{x}}{\partial u^n}$
the covariant components of the metric tensor, $A_k$ the covariant
components of the vector potential and $\Phi_E$ the electrostatic
potential (if any).
The Euler-Lagrange equations $\frac{d}{dt}\left(\frac{\partial
    \mathcal{L}}{\partial v^j}\right) = \frac{\partial
  \mathcal{L}}{\partial u^j}$ eventually yield the following equations
of motion in the context of time-independent coordinate systems,
\begin{align}
\label{eqn:lorentz_eom}
\dot{v}^i + v^mv^n \Gamma^i_{mn}  &= \frac{q}{m} g^{ij} \sqrt{g}\epsilon_{jkl} v^k B^l +
\frac{q}{m} E^i
\end{align}
where $\epsilon_{ijk}$ is the anti-symmetric Levi-Cività tensor,
$\Gamma^i_{mn}=\frac{1}{2}g^{ij}\left(\partial_m g_{nj} + \partial_n
  g_{jm} - \partial_j g_{mn}\right)$ the Christoffel symbol of second
type compensating for inertial forces due to the curvilinear
coordinate system, $B^i$ the contravariant components of the magnetic
field and $E^i$ the contravariant components of the electric field. In
the framework of general relativity, this result is identical to the
geodesic equations of a charged particle.

\begin{figure}[!]
  \includegraphics[width=0.39\linewidth]{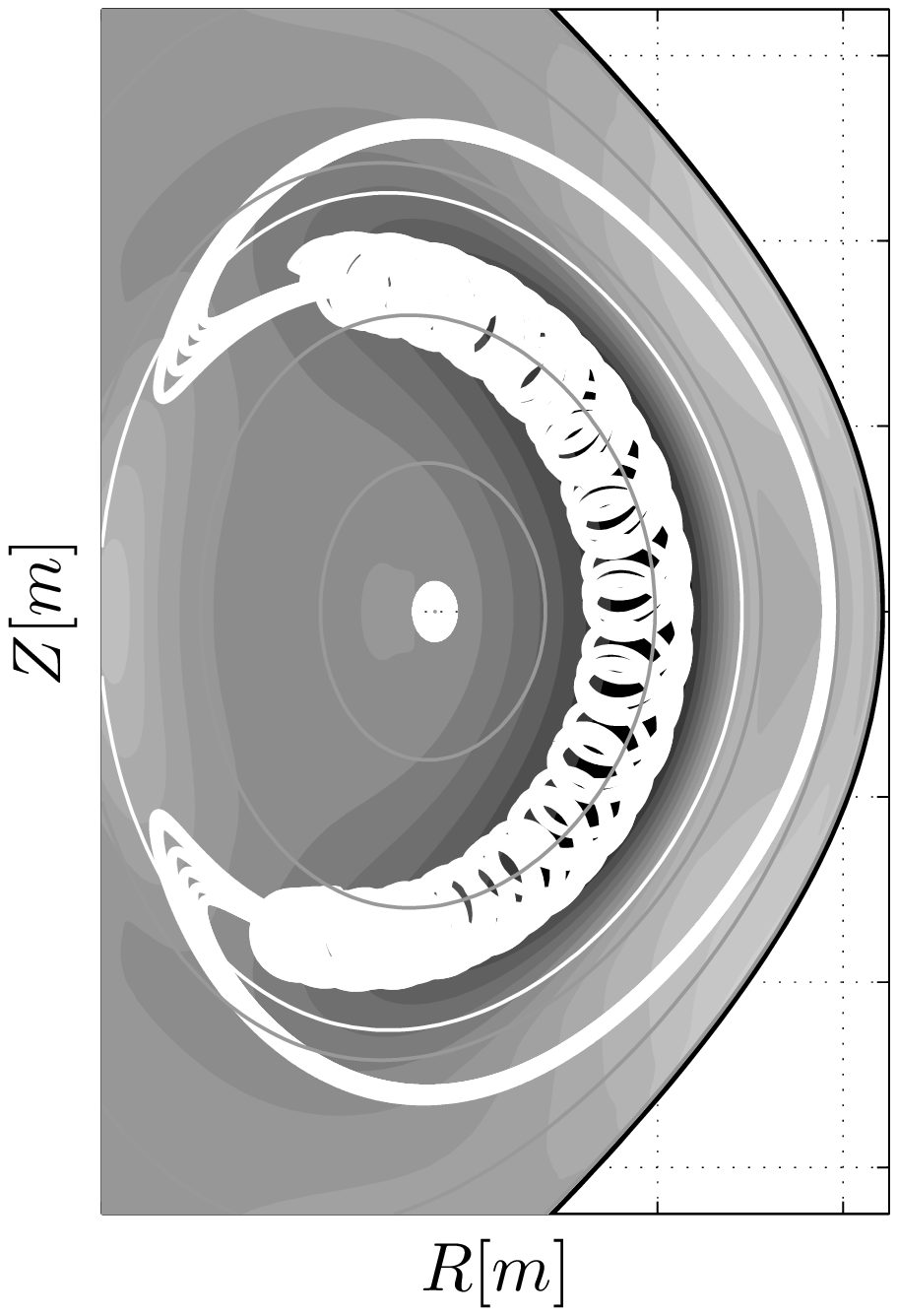}
  \includegraphics[width=0.59\linewidth]{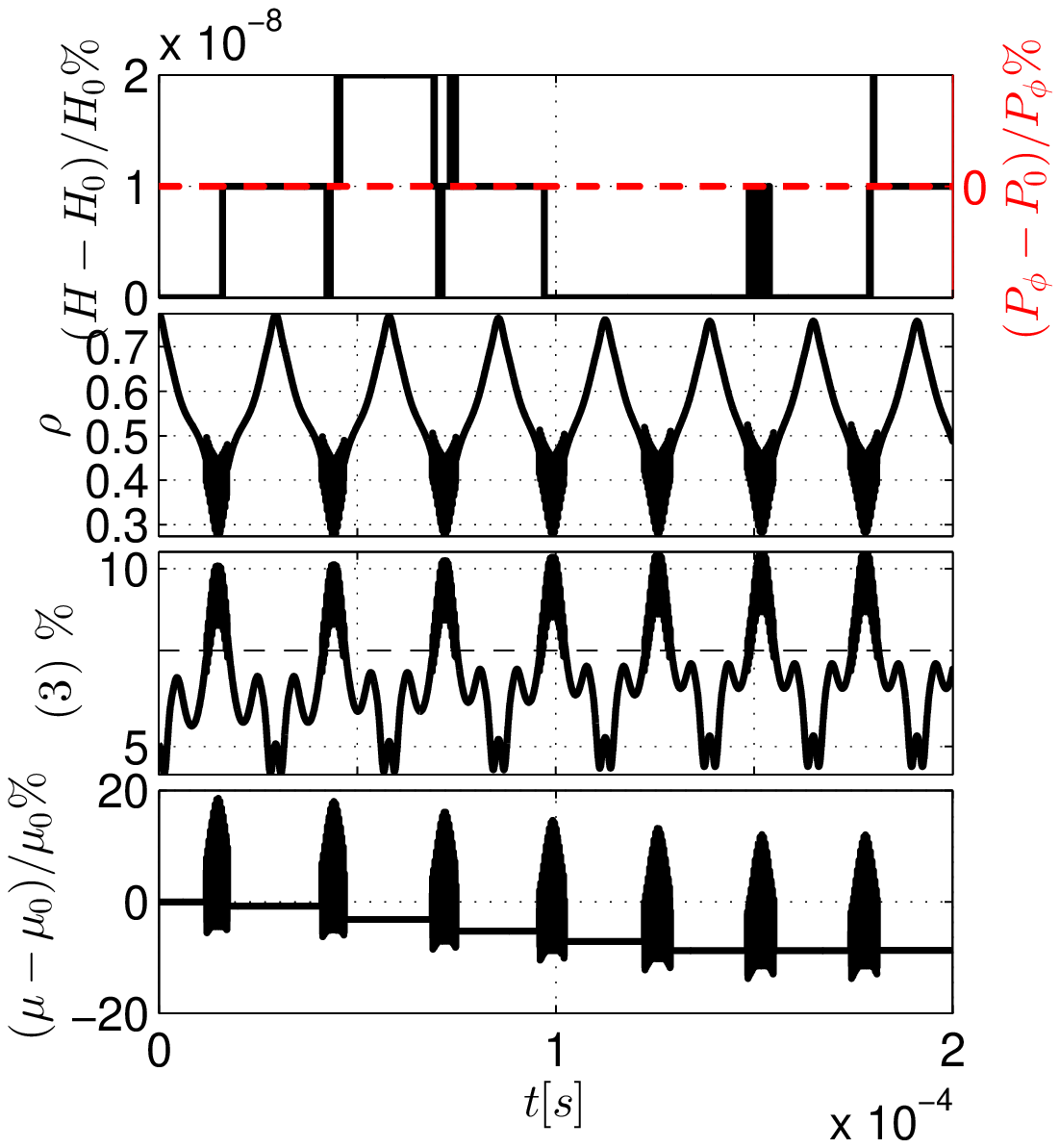}
  \caption{\label{fig:switching} Illustration of the
      switching algorithm in a MAST axisymmetric hybrid plasma.\\
      (left) banana orbit of a $D^+$ ion at $\mathcal{H}=10$ KeV,
      $v_{||0}/v=0.72$ and $\rho_0=0.77$ computed over 7 bounce
      periods ($2\times 10^{-4}$s) using GCDE/full-orbit switching
      with a threshold for criterion (\ref{eqn:gc_condition}) of
      $7.3\%$. The background colours highlight the field variation
      felt by the particle (dark black patch is above the
      threshold). Light grey lines depict flux surfaces.\\
      (right) the first plot shows the relative error in energy (solid
      black line) and relative error in toroidal momentum (dashed red
      line) in time. The invariance of $\mathcal{H}$ and $P_\phi$
      converges to machine precision ($\sim 10^{-8}\%$) for this
      collisionless orbit by using a sufficiently small time-step. The
      second plot is the evolution of the particle's radial coordinate
      $\rho=\sqrt{\Phi/\Phi_e}$, where $\Phi$ is the poloidal magnetic
      flux ($\Phi_e$ its value on the last closed flux-surface). The
      third plot displays the field variation from equation
      (\ref{eqn:gc_condition}) felt at the guiding-centre position
      (black). The last plot displays the relative variation of the
      first-order magnetic moment $\mu = m v_\perp^2/2B$ in time.}
\end{figure}

Solving the full-orbit equations is computationally heavy. Worse, the
smaller the Larmor radius, the harder it is to resolve the small
gyration of the particle around the field-lines and a full-orbit
calculation becomes less precise than using a guiding-centre
formulation. It is useful to be able to switch instantaneously between
GCDE and full-orbit equations and to focus numerical resources only
where the GCDE are inadequate. Figure \ref{fig:switching} shows an
example of a $10$ KeV $D^+$ ion undergoing a banana orbit in an
axisymmetric equilibrium of a MAST hybrid plasma. When its
guiding-centre enters an area of the magnetic configuration where the
field variation criterion (\ref{eqn:gc_condition}) is above the
arbitrary threshold choice of $7.3\%$, the algorithm switches between
GCDE and full-orbit equations and vice-versa when
(\ref{eqn:gc_condition}) is below $7.3\%$.

The mapping between particle and guiding-centre variables is performed
here at first order in a $\rho_L$ expansion. This implies that the
guiding-centre trajectory follows the average particle motion with an
error that scales proportionally to the field variation
$|(\bm{\rho}_L\cdot\grad) \bm{B}|/B$, as explained in section
\ref{sec:field_variation}. Divergences are minimised by making sure
that constants of motion match in the two phase-spaces and that the
switching process is triggered in an area of the magnetic field where
its variation is small. A small discrepancy is yet induced when
switching from guiding-centre to particle variables, as observed on
the left plot of figure \ref{fig:switching}. In this example of a
barely trapped particle, despite exact conservation of energy and
toroidal momentum, a slight change in the position of the bounce tips
is caused by the imperfect match of $\mu$ at first-order in Larmor
radius between the particle and the guiding-centre. The magnetic
moment $\mu$ varies by $20\%$ during the full-orbit portion. A
second-order expansion of GCDE as in \cite{belova} would help to
reduce this discrepancy.

\begin{figure}[!]
  \includegraphics[width=0.39\linewidth]{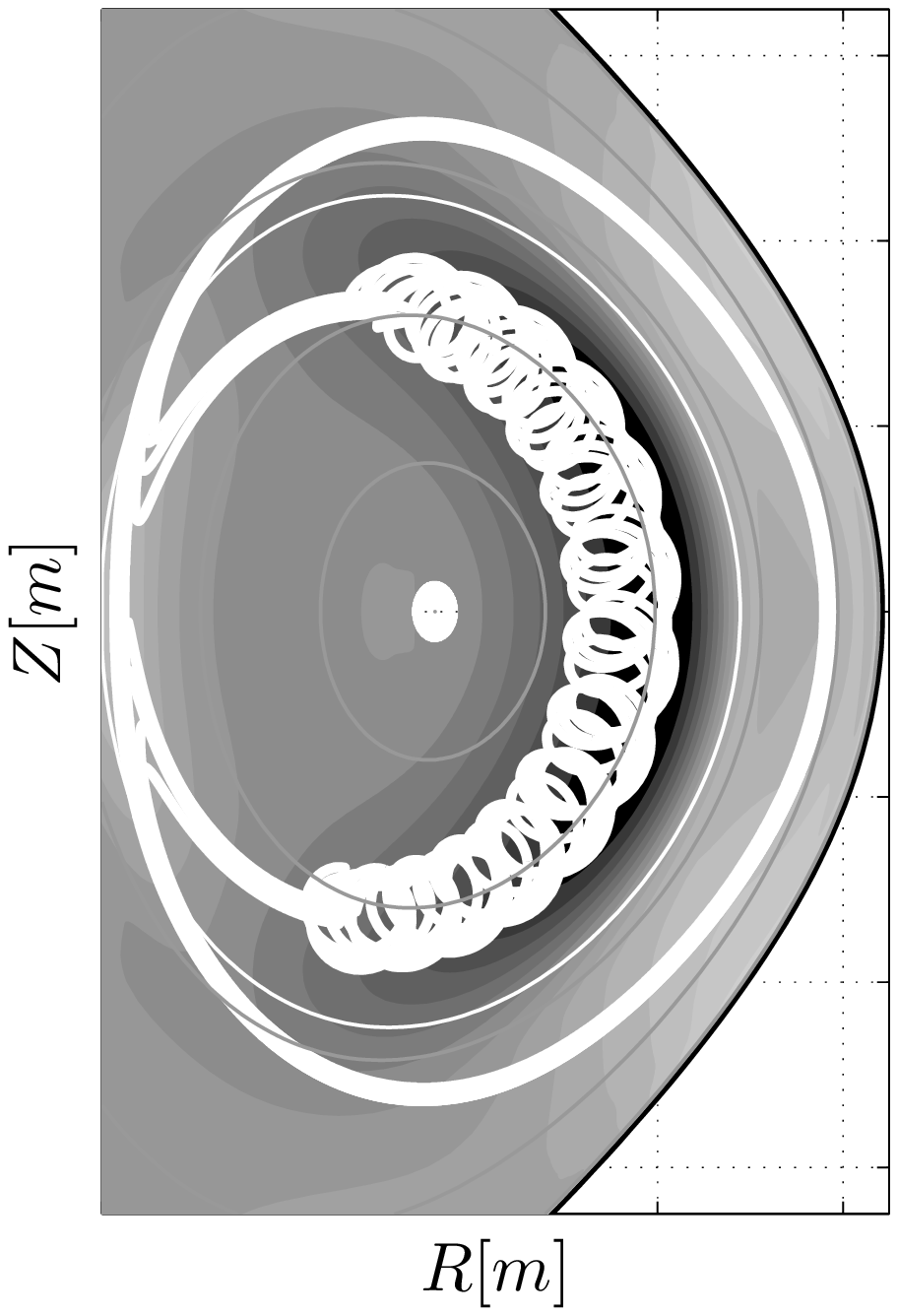}
  \includegraphics[width=0.59\linewidth]{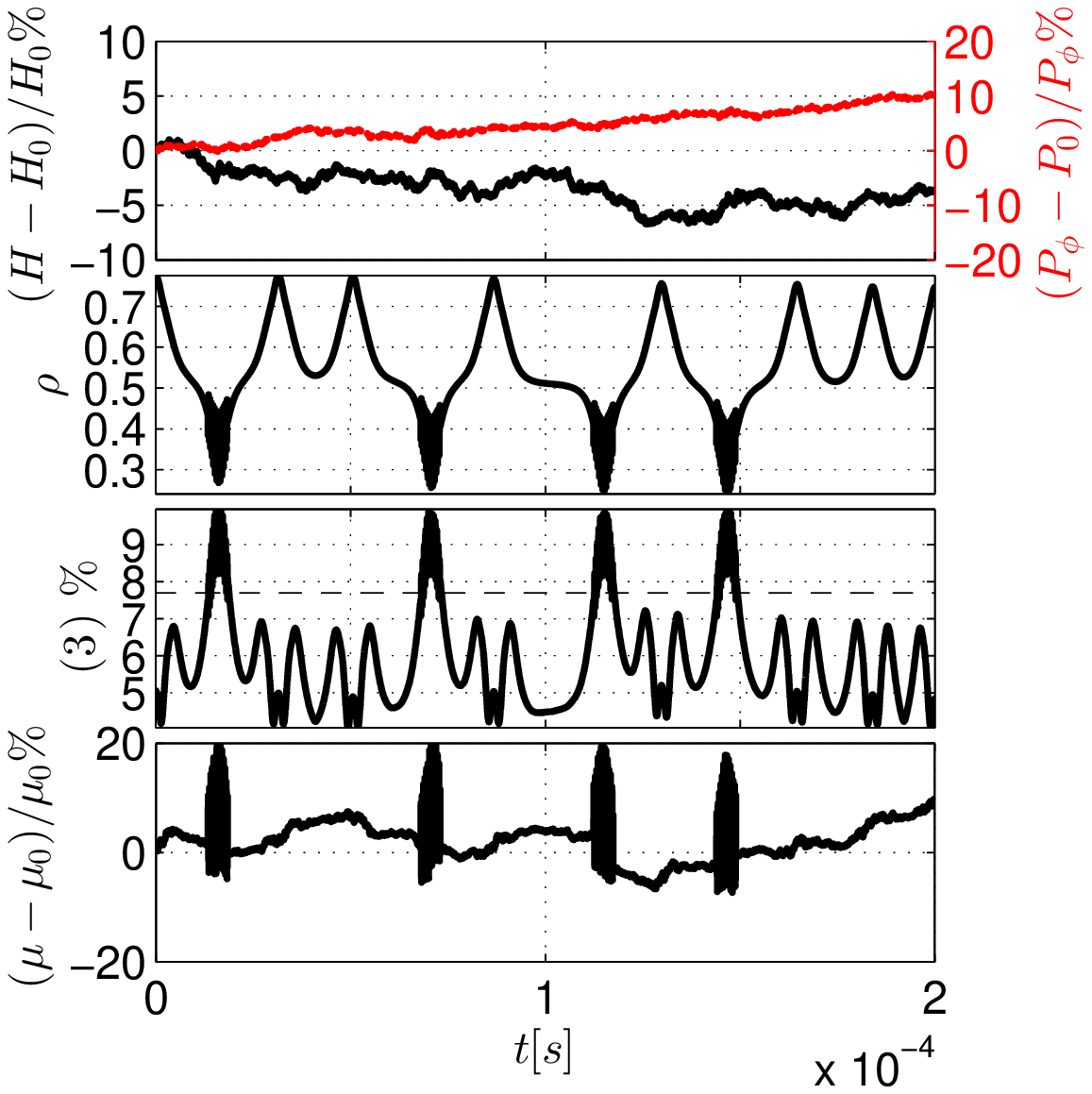}
  \caption{\label{fig:switching_collisions} Orbit of the same $D^+$
    particle as in figure \ref{fig:switching} ($\mathcal{H}_0=10$keV),
    perturbed by the action of Monte-Carlo operators simulating
    Coulomb collisions with the background plasma
    \cite{boozer-kuopetravic,albergante-2012}. The particle receives
    random kicks in energy $\mathcal{H}$ and pitch-variable $v_{||}/v$
    which affect the position of the bounce tips and the constants of
    motion to greater extent than the switching between guiding-centre
    and particle phase-space. Energy $\mathcal{H}$, toroidal momentum
    $P_\phi$ and magnetic moment $\mu$ are varied by $5-10\%$ percent
    over $2\times 10^{-4}$s, consistent with a scattering time of
    $\sim 2.4\times 10^{-2}$s (background electron temperature of
    $T_e\approx 1$keV, electron density of
    $n_e=3\times 10^{19}m^{-3}$).}
\end{figure}

This switching algorithm is appropriate in conjunction with
Monte-Carlo operators (for e.g. Coulomb collisions with the
background, charge exchange, anomalous transport, etc.), where the
random kicks perturb the particle's motion to greater extent than
the switching process, as observed in figure
\ref{fig:switching_collisions}.
\subsubsection{From particle to guiding-centre variables}
The mapping from the particle coordinates $(\bm{x},\bm{v})$ to the
guiding-centre coordinates $(\bm{X},v_{||},\mu)$ is a surjection in
the sense that an infinite number of particles share the same
guiding-centre. At first order, the guiding-centre position is
calculated from the particle variables as
\begin{equation}
\label{eqn:gcpos-from-fl}
  \bm{X} = \bm{x} - \bm{\rho}_L(\bm{x}) = \bm{x} + \frac{m}{q B(\bm{x})}\bm{v}\times\bm{b}(\bm{x}).
\end{equation}
It is recommended to do this displacement in Cartesian coordinates if
the transformation from Cartesian to curvilinear coordinates is known
(or if an inverse mapping or fast root finding algorithm is
available). Otherwise, it helps to perform this operation in a
pseudo-Cartesian coordinate system where the basis vectors are nearly
unit vectors, and then invert back to curvilinear coordinates (see
appendix \ref{sec:curvcoord}).

After setting the guiding-centre position, it is important to ensure
that the guiding-centre energy and the particle energy are equal since
they are identical constants of motion in the presence of static
electromagnetic fields, i.e.
\begin{align*}
  \mathcal{H} &= \frac{1}{2}m v^2 + q\Phi_E(\bm{x}) = 
  \frac{1}{2}m g_{ij}(\bm{x})v^i v^j + q\Phi_E(\bm{x}) \\
  &= \frac{1}{2} m v_{||}^2 + \mu B(\bm{X}) + q\Phi_E(\bm{X}).
\end{align*}
The particle toroidal momentum and the guiding-centre's in
axisymmetric systems are also supposed to match, i.e.
\begin{align*}
  P_\phi &= qA_\phi + m v_\phi = qA_\phi(\bm{x}) + m g_{i\phi}(\bm{x}) v^i \\
  &= qA_\phi(\bm{X}) + m v_{||} b_\phi(\bm{X}).
\end{align*}
From the two previous equations, even for non-axisymmetric or
time-varying fields, it is reasonable to express the guiding-centre
velocity variables as a function of the particle phase-space
coordinates in the following way
\begin{align}
  \label{eqn:gcvpar-from-fl}
  v_{||} &= \frac{B(\bm{X})}{B_\phi(\bm{X})}\left\{g_{i\phi}(\bm{x}) v^i+\frac{q}{m}\left[ A_\phi(\bm{x})-A_\phi(\bm{X})\right]\right\}\\
  \label{eqn:gcmu-from-fl}
  \mu &= \frac{1}{2}\frac{m}{B(\bm{X})} [g_{ij}(\bm{x})v^i v^j - v_{||}^2] + \frac{q}{B(\bm{X})}[\Phi_E(\bm{x})-\Phi_E(\bm{X})].
\end{align}
By doing so, neither the energy nor the toroidal momentum of the
particle can be seen to change on figure \ref{fig:switching}.

In magnetic field representations relevant for tokamak and stellarator
equilibria, the toroidal component of the vector potential is, up to a
gauge choice, equal to the poloidal magnetic flux $A_\phi=-\Psi(\rho)$ (an
important flux function sometimes used as a radial
coordinate). Equations
(\ref{eqn:gcvpar-from-fl}-\ref{eqn:gcmu-from-fl}) are simplified by
linearly expanding in gyro-radius, for example
$A_\phi(\bm{x})-A_\phi(\bm{X}) \approx -\bm{\rho}_L \cdot\grad \Psi$
and $\Phi_E(\bm{x})-\Phi_E(\bm{X}) \approx \bm{\rho}_L\cdot \grad
\Phi_E = -\bm{\rho}_L\cdot \bm{E}$.
\subsubsection{From guiding-centre to particle variables}
Changing from the guiding-centre to the particle phase-space is not as
straight-forward, in the sense that the direction of $\bm{\rho}_L$ in
the perpendicular plane to $\bm{b}$ can be chosen arbitrarily. For
most plasma relevant magnetic fields, the dominant source of field
variation appears from the gradient of the field strength. The
gyro-angle is thus chosen such that the modulus of the magnetic field
varies the least from the particle position to the guiding-centre
position. The norm of the perpendicular velocity at both positions is
then nearly the same
\begin{align*}
  v_\perp^2(\bm{X}) = \frac{2\mu}{m} B(\bm{X}) \approx \frac{2\mu}{m}B(\bm{x}) = v_\perp^2(\bm{x})
\end{align*}
The same argument applied to $\mathcal{H} = \frac{1}{2}mv_{||}^2 +\mu B + q
\Phi_E$ shows that the parallel velocity $v_{||}^2(\bm{X})\approx
v_{||}^2(\bm{x}) + q[\Phi_E(\bm{x})-\Phi_E(\bm{X})]/2m$. To decide the
particle position, the Larmor radius vector $\bm{\rho}_L$ is thus
chosen perpendicularly to $\grad B$
\begin{equation}
  \bm{x} = \bm{X} + \bm{\rho}_L(\bm{X}) = \bm{X} \pm \left.\rho_L\frac{\bm{B}\times\grad
      B}{\left|\bm{B}\times\grad B\right|} \right|_{\bm{X}}
  \label{eqn:larmor_prescription}
\end{equation}
where $\rho_L =
\sqrt{\frac{2}{B(\bm{X})}\frac{m}{q}\frac{\mu}{q}}\approx
\sqrt{\frac{2}{B(\bm{x})}\frac{m}{q}\frac{\mu}{q}} $. With this
prescription, the perpendicular velocity vector at the guiding-centre
position is determined as
\begin{align*}
\bm{v}_\perp(\bm{X}) = \frac{q}{m}\bm{\rho}_L\times\bm{B}(\bm{X})
= v_\perp \left.\frac{\grad B  - \bm{b}(\bm{b}\cdot\grad B)}{\left|\bm{B}\times\grad B\right|}\right|_{\bm{X}}
\end{align*}
The total velocity vector of the particle at the guiding-centre
position is
\begin{equation*}
  \bm{v}(\bm{X}) = v_{||}(\bm{X}) \bm{b}(\bm{X}) + \bm{v}_\perp(\bm{X}) \overset{!}{=} \bm{v}(\bm{x})
\end{equation*}
and is assumed to be the same at the particle position. In Cartesian
coordinates, this means that the components of the velocity vector are
equal at both locations. In curvilinear coordinates, these components
must be re-evaluated because the basis vectors change (mathematical
problem referred to as \emph{parallel transport} of a vector)
\begin{align}
  v^i(\bm{x}) &= \bm{v}(\bm{X})\cdot \grad u^i(\bm{x}) =
  v^j(\bm{X})\bm{e}_j(\bm{X})\cdot\grad u^i(\bm{x}) \nonumber\\
  &=\frac{\partial u^i}{\partial x^k}\Bigg|_{\bm{x}} \frac{\partial x^k}{\partial u^j}\Bigg|_{\bm{X}} v^j
  = [\Lambda^{-T}(\bm{x}) \Lambda(\bm{X}) ]^i_{\ j} v^j(\bm{X}).
  \label{eqn:init_velo}
\end{align}
This operation preserves the modulus of the velocity vector,
$v^2=g_{ij}(\bm{x}) v^i(\bm{x})v^j(\bm{x})= g_{ij}(\bm{X}) v^i(\bm{X})
v^j(\bm{X})$,
and ensures that the particle's energy is defined equally to the
guiding-centre's. Applying several coordinate transformations (matrix
multiplications) is however sensitive to truncation error, as observed
from the relative increase in energy in the bottom plot of figure
\ref{fig:switching}.

In axisymmetry, matching $P_\phi$ from the guiding-centre to the
particle position is enforced so that there is no variation of the red
dashed curve in first plot of \ref{fig:switching}. This is performed
by scaling the parallel and perpendicular component
accordingly. Depending on the field variation, a small discrepancy is
induced by the variation of the magnetic moment $\mu$ along the
full-orbit. This produces a small change in the bounce tips position
which eventually saturates over successive switches after an
artificial re-positioning. Barely trapped particles are most sensitive
to changes in $\mu$, so figure \ref{fig:switching} illustrates a
delicate example. A better match is obtained between guiding-centre
and full-orbits of passing particles. The discrepancy due to the
switching is easily controlled by adjusting the threshold of criterion
(\ref{eqn:gc_condition}). The action of Monte-Carlo operators in
slowing-down simulations has a more important impact on the orbits, as
observed on figure \ref{fig:switching_collisions} where collisions
with the background plasma causes de-trapping and diffusion.


%
\section{Fast ion orbits in MAST helical core}
\label{sec:heli_analysis}
The MAST Long-Lived Mode (LLM) \cite{chapman-2010} is assumed to be a
saturated ideal internal kink mode with dominant non-axisymmetric
$n=1$ mode structure, that can be modelled as a 3D helical core
magnetic equilibrium \cite{cooper-helicalmast} using codes such as
\code{VMEC} \cite{vmec} or \code{ANIMEC} \cite{animec}. This static
solution to the ideal MHD force balance equation is found for slightly
reversed q-profile\footnote{In this section, the symbol $q$ will refer
  to the q-profile and should not be confused with its earlier use as
  the particle charge.} where $q_{min}\sim 1$ (but $q>1$). The
amplitude of the helical displacement normalised to the minor radius,
noted $\delta_h =\sqrt{R_{01}(0)^2+Z_{01}(0)^2}/a$, probably reaches
more than $0.1$ in MAST \cite{chapman-2010,brunetti-2014}. The
confinement of fast ions is observed experimentally to be affected
during LLM activity in MAST, as particles are relocated away from the
central region of the plasma \cite{cecconello-2012}.  Numerical
studies using guiding-centre theory have modelled the redistribution
of NBI fast ions during these helical states \cite{pfefferle-nf}. In
what follows, criterion (\ref{eqn:gc_condition}) is applied to fast
ions in an extreme helical core case where $\delta_h=0.23$, in order
to visualise the regions where the field variation is
severe. Full-orbits are verified to reproduce the exotic
guiding-centre drift motion that was encountered in previous work.
\subsubsection{Magnetic field variations}
\label{sec:heli_fieldvariation}
\begin{figure}
  \center\includegraphics[width=\linewidth]{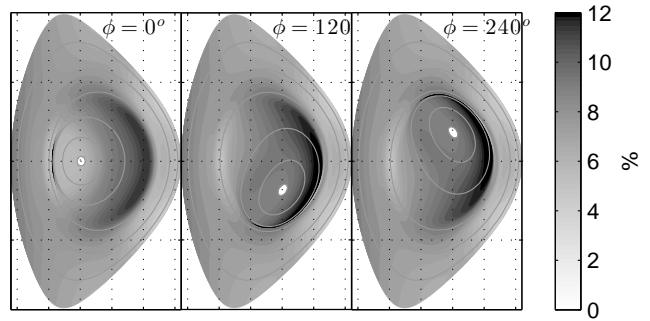}
  \caption{\label{fig:fieldvariation} Field variation criterion
    (\ref{eqn:gc_condition}) represented on $R-Z$ planes at successive
    toroidal angles and applied to MAST helical core of
    $\delta_h=0.23$ and a $D^+$ ion with $\mathcal{H}_\perp=10$ KeV, $\rho_L
    \sim 5.5$ cm. Gray dashed lines depict flux-surfaces; at $\phi=0$,
    the compressed region is around $\theta=\pi$ and uncompressed
    around $\theta=0$.}
\end{figure}
\begin{figure}
  \center\includegraphics[width=\linewidth]{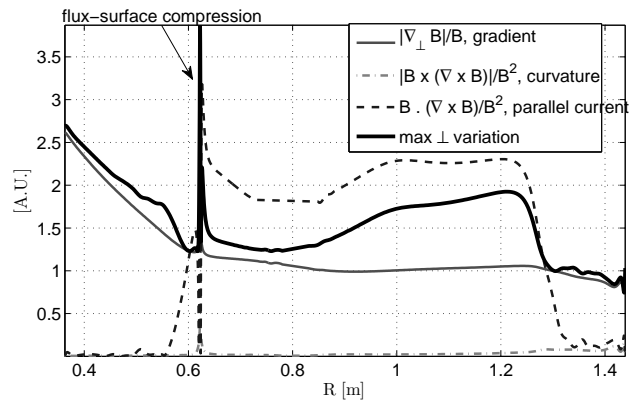}
  \caption{\label{fig:fv_decomposition} Profile of the maximum field
    variation along the mid-plane at $\phi=0$. Comparison with
    gradient, curvature and parallel current terms that contribute in
    the total variation of the field. The field variation is mostly
    due to the gradient of $|B|$, except in the low-field side of the
    core and the region of compressed flux-surfaces where the parallel
    curl becomes large.}
\end{figure}
\begin{figure}
  \subfigure[\label{fig:heli_fieldlines} Path of a single field-line
  on the $\theta-\phi$ plane (geometric angles) at $\rho=r/a = 0.48$
  (transition region).]%
  {\includegraphics[width=0.42\linewidth]{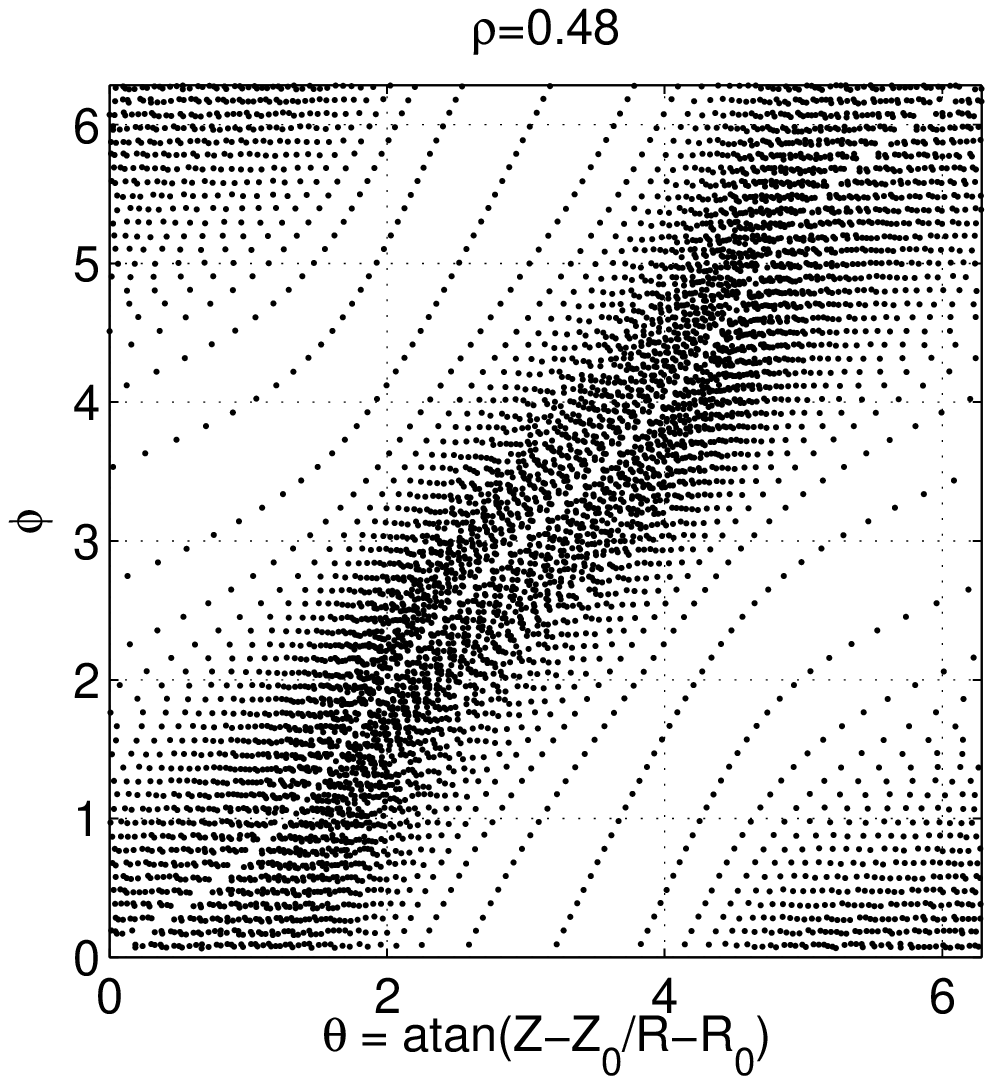}}
  \subfigure[\label{fig:heli_paracurrent} Parallel current as a
  function of flux label]%
  {\includegraphics[width=0.55\linewidth]{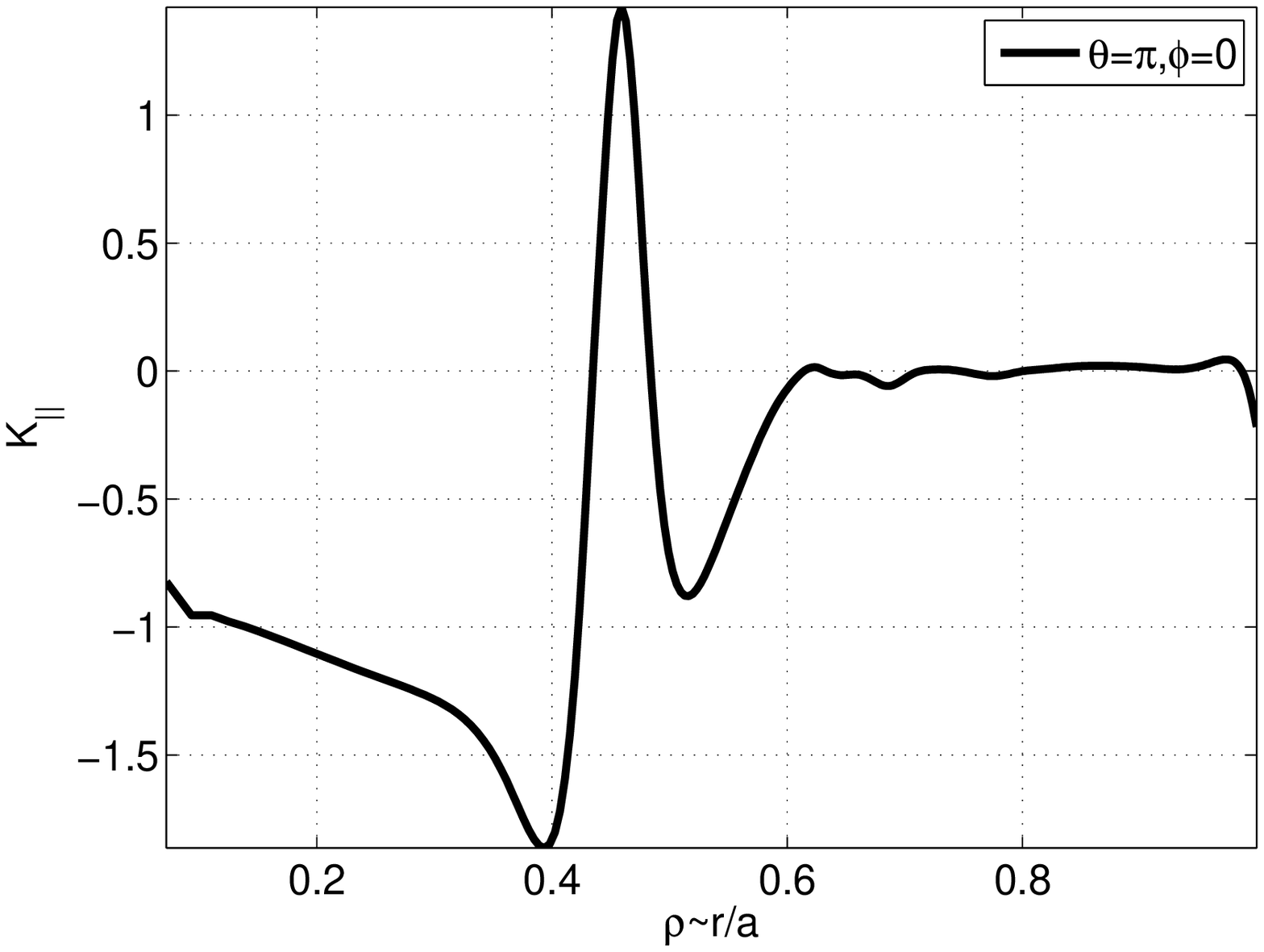}}\\
  \caption{\label{heli_characteristics} Characteristics of the
    transition region between the helical core and the axisymmetric
    mantle of the MAST $\delta_h=0.23$ distorted equilibrium.}
\end{figure}

Due to the low-aspect ratio of the MAST tokamak and its relatively
high value of plasma beta, the field variation can become quite strong
even in the case of axisymmetry. The widest problematic region for
first-order GCDE is located on the low field side of the core where
the field variation can rise up to $12\%$ for deuterium ions with
$\mathcal{H}_\perp=10$ KeV ($\rho_L\sim 5$ cm), as seen on figure
\ref{fig:fieldvariation} showing a map of the field variation at
different toroidal angles for an equilibrium with a helical core of
amplitude $\delta_h = 0.23$. Strong parallel currents develop in the
transition region between the helical core and the axisymmetric mantle
(see figure \ref{fig:heli_paracurrent}) because the q-profile is close
to unity. Figure \ref{fig:fv_decomposition} shows the field variation
profile at $\phi=0$ and at $Z=0$ (mid-plane) as a function of major
radius as well as its various constituents. It is observed that the
bump on the low-field side of the core is mainly caused by the
parallel curl of the magnetic field. The discrepancies between
guiding-centre and full-orbit calculations are expected to be
magnified in this region, mostly affecting the interpretation of the
parallel velocity, as discussed at the beginning of section
\ref{sec:field_variation} and in appendix
\ref{sec:orbit_sheared_field}.

Due to the geometry of the helical kink, flux-surfaces are tightly
compressed against the axisymmetric mantle on one side of the core,
whereas they separate on the other side (see light grey lines
representing flux-surfaces in figure
\ref{fig:fieldvariation}). Because of the abrupt transition between
helical core and axisymmetric mantle, the field variation spikes in
the zone where the flux-surfaces are compressed, as seen on figure
\ref{fig:fv_decomposition}. The field-lines effectively avoid this
zone, as demonstrated in figure \ref{fig:heli_fieldlines} displaying
the path of a single field-line at a given flux label in the
transition region on the periodic plane $(\theta,\phi)$ formed by the
geometric poloidal and toroidal angles. The $\phi-\theta=0$
uncompressed region is more densely visited by field-lines than the
$\phi-\theta=\pi$ compressed region such that their bending (S-shape)
is more pronounced in the latter region. Even though the shear of the
q-profile is low (zero at $q_{min}$), the local pitch of the field
varies extensively within a flux-surface as well as from one surface
to the next, such that the field-lines are locally strongly sheared,
analogously to example (\ref{eqn:sheared_magfield}). In the limit
where the $q$-profile reaches unity and in the framework of ideal MHD,
an infinite current sheet will form. Resistive effects would however
soften the sharp transition in generating magnetic islands and a
different approach for generating the equilibrium should be considered
for the resonant $q$ case.
The smooth representation of the abrupt transition between helical
core and axisymmetric mantle was greatly facilitated by the
Spline-Fourier interpolation scheme \cite{pfefferle-cpc}.
\subsubsection{Full-orbits and guiding-centre drift surfaces}
\label{sec:heli_orbits}
\begin{figure}
  \centering \subfigure[\label{fig:helicalcore_contours} Particles
  staying in the helical region draw out drift surfaces, projected as
  closed contours on poloidal planes.]%
  {\includegraphics[width=\columnwidth]{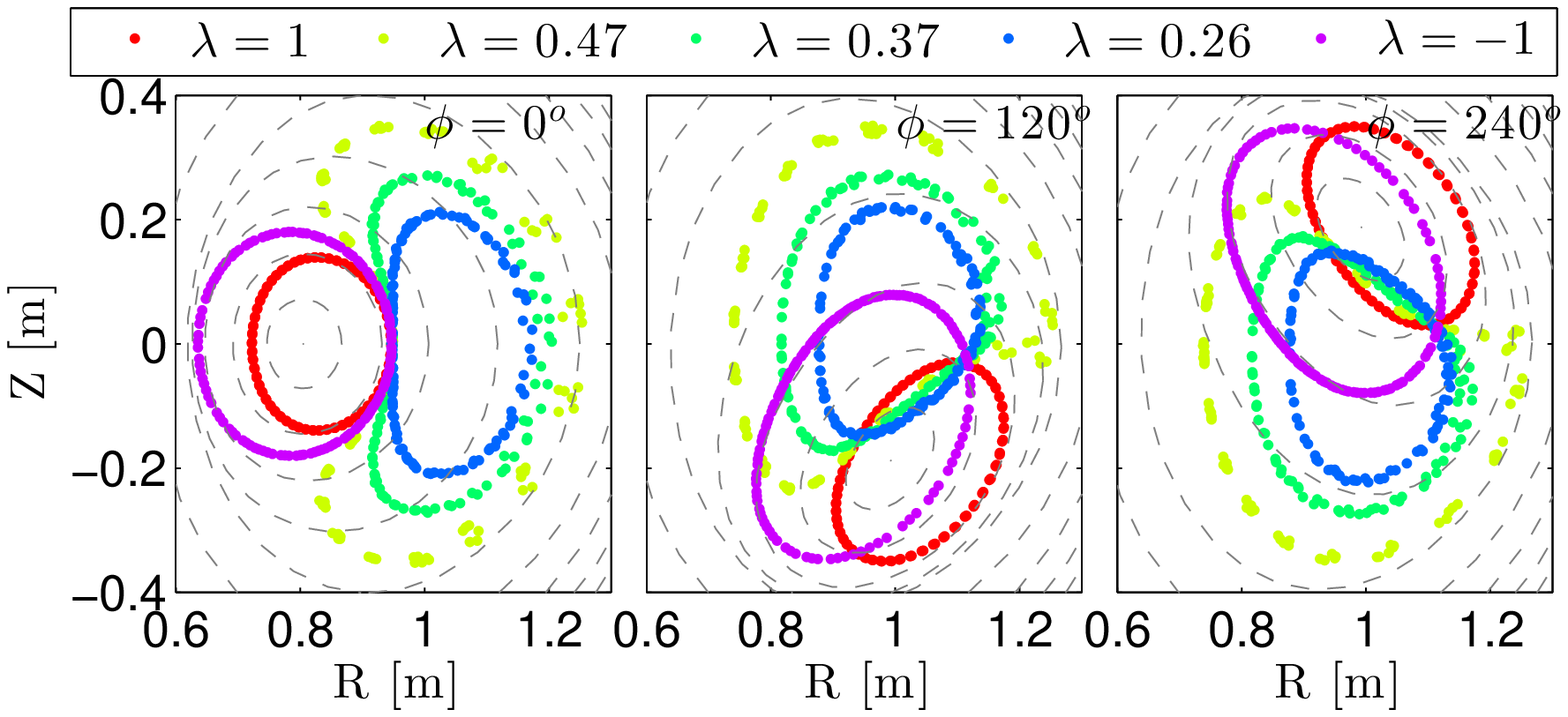}}
  \subfigure[\label{fig:chaoticdrift} Particles transitioning between
  helical core and axisymmetric mantle display chaotic motion.]%
  {\includegraphics[width=\columnwidth]{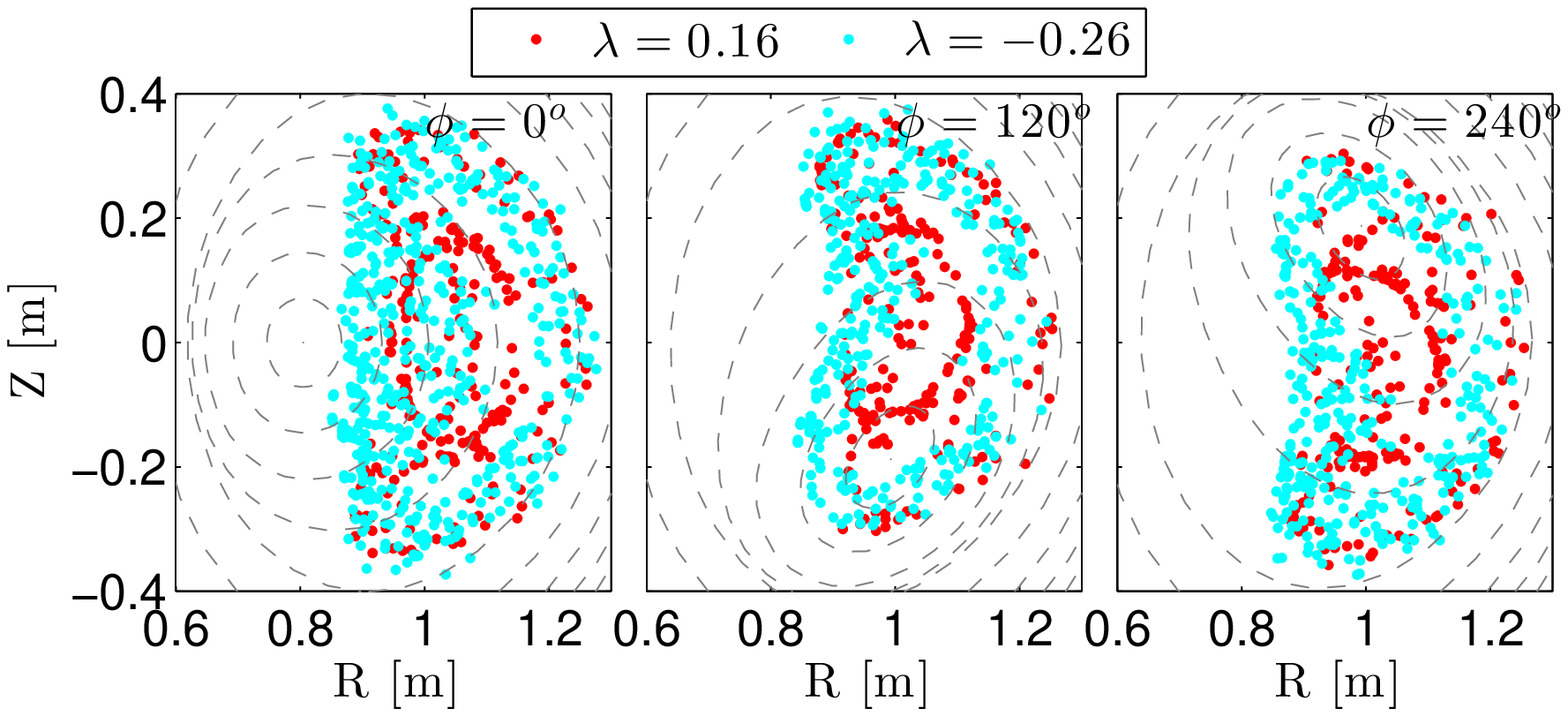}}
  \caption{Poincaré section of the guiding-centre trajectories of a
    set of 10 KeV $D^+$ ions initialised at
    $\rho=\sqrt{\Phi/\Phi_e}=0.22$ and $\phi=\theta=0$ but with
    different pitch-variable $\lambda=v_{||}/v$. The guiding-centre
    position are traced from the particles' full-orbit in this MAST
    $\delta_h=0.23$ helical core.}
\label{fig:lambdascan_cuts}
\end{figure}

In the helical core equilibrium just as in axisymmetric magnetic
fields, the guiding-centre trajectories of energetic particles sketch
out \emph{drift surfaces}. In axisymmetric magnetic fields, drift
surfaces are a consequence of the conservation of toroidal momentum,
effectively guaranteeing particle confinement. In helical states, the
fact that drift surfaces exist indicates that there is a non-trivial
constant of motion (possibly a combination of toroidal and poloidal
momentum\footnote{Typically, in the uncompressed region, the
  $q$-profile is close to unity and relatively constant (low shear),
  which would indicate helical symmetry.}). On a poloidal cut at a
fixed toroidal angle, drift surfaces are reduced to closed contours,
progressively drawn by the punctures of the guiding-centre trajectory
across that same vertical plane. Drift surfaces are present as long as
particles do not transit from the helical region to the axisymmetric
outer mantle, as seen figure \ref{fig:helicalcore_contours} showing
the guiding-centre trajectories computed from full particle orbits. If
they exit the core, their motion is randomised by successively
following helical and axisymmetric drift surfaces (see figure
\ref{fig:chaoticdrift}). These examples show that particle confinement
is not necessarily degraded by the helical kink, moreover that the
outer mantle is acting as a confining pinch.

The magnetic axis naturally intersects some drift surfaces located in
the core region such that orbits often pass through that singular
point. This is a major difference with axisymmetric cases where
difficulties with the magnetic axis only occur through collisional
processes.
\subsubsection{Deeply-passing particles}
\begin{figure}
  \centering \subfigure[\label{fig:poincare-co} Co-passing particles]%
  {\includegraphics[width=\columnwidth]{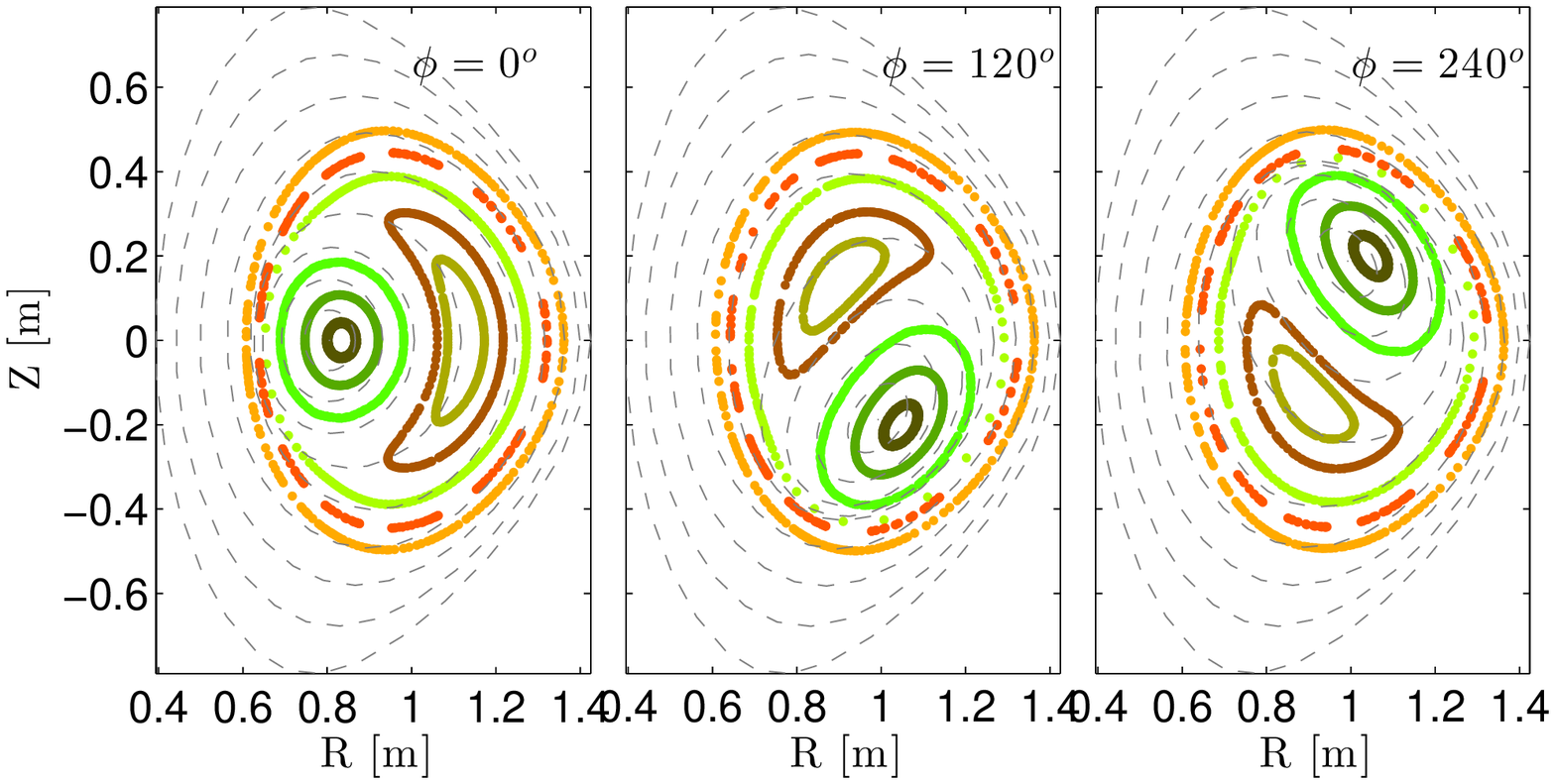}}
  \subfigure[\label{fig:poincare-ct} Counter-passing particles]%
  {\includegraphics[width=\columnwidth]{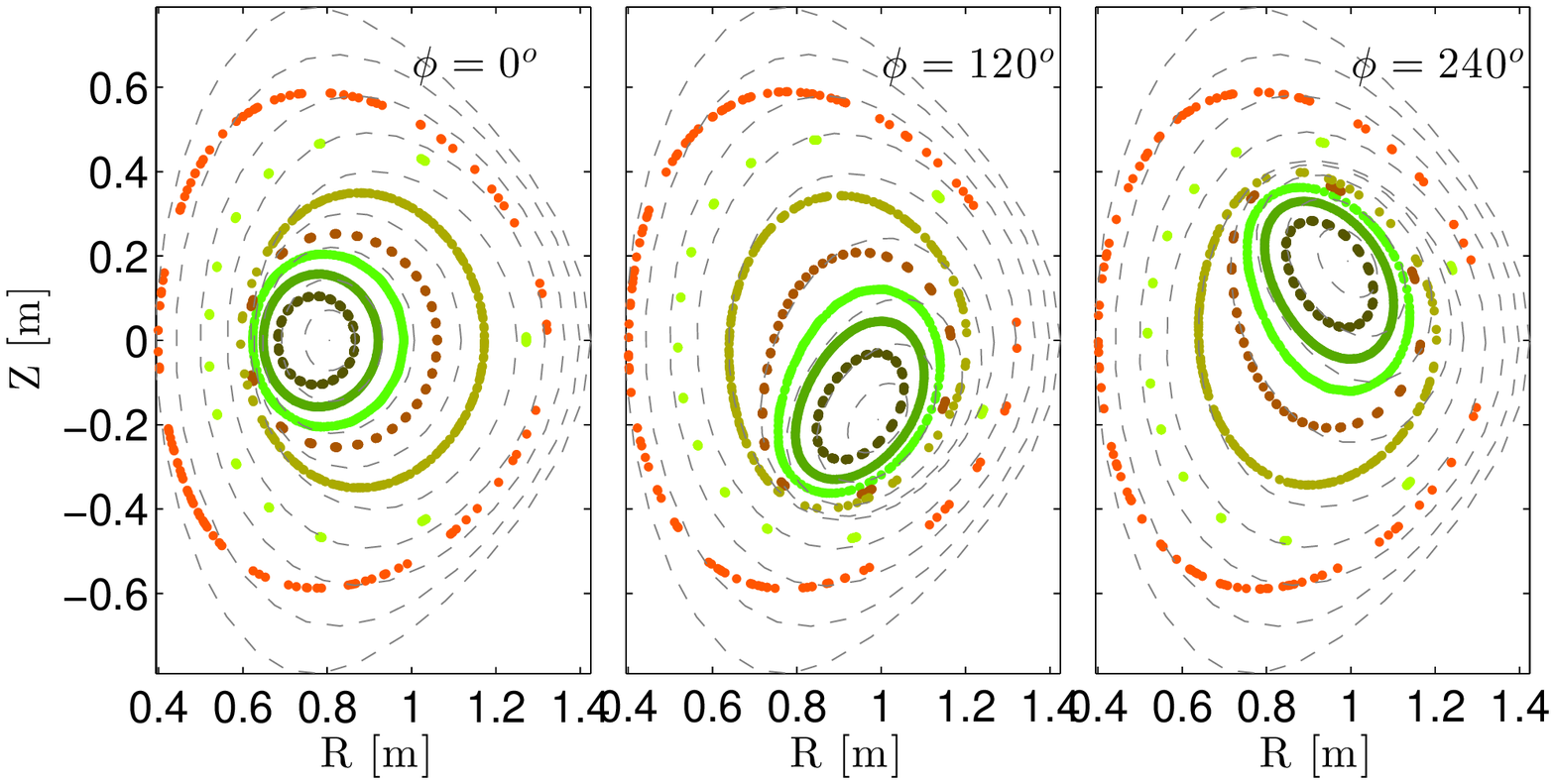}}
  \caption{Poincaré cut of drift surfaces created by $10$ KeV $D^+$
    ions initially with $v_{||}/v=\pm 1$ as computed using full-orbit
    equations in MAST helical core with $\delta_h=0.23$ displacement.}
\label{fig:poincare}
\end{figure}

The MAST Positive Ion Neutral Injector (PINI) emits neutrals at the
energy of $60$ KeV. Larmor radii of trapped particles at these
energies are a few centimetres large, but since injection is
tangential, the kind of fast particles produced are almost exclusively
deeply-passing particles, especially in the core region near the
helical magnetic axis. It is instructive to focus on single particle
orbits with initial pitch-variable $\lambda=v_{||}/v$ close to unity
(small magnetic moment). In particular, the guiding-centre equations
of particles with a pitch-variable equal to unity (zero magnetic
moment), yield $v_{||}=v$ constant. For such value of the
pitch-variable, the drift surfaces associated to their motion are
closest to flux-surfaces, as seen on figure
\ref{fig:lambdascan_cuts}. While all counter-passing particles travel
around the original magnetic axis (see figure \ref{fig:poincare-ct}),
some co-passing particles circulate on drift surfaces that do not
enclose the helical magnetic axis in the region of uncompressed
flux-surfaces (see figure \ref{fig:poincare-co}). They move around a
\emph{helical drift axis}, positioned on the opposite side of the
magnetic axis inside the kinked core. That behaviour is in part
responsible for off-axis redistribution of NBI fast particles,
evidenced by MAST neutron camera traces \cite{cecconello-2012} and
illustrated using a guiding-centre drift approach in slowing-down
simulations \cite{pfefferle-nf}.
In a full-orbit formulation, particles can develop small gyro-motion
even if their pitch-variable is initially equal to unity (zero Larmor
radius) depending on the bending of field-lines. The magnetic moment
$\mu = m v_\perp^2/2B$ is not constant, but oscillates around an
average value. In the case of helical cores, it is found that, even
with strong field variations in the core region, full-orbit
calculations of deeply-passing particles yield similar drift
surfaces. Baños corrections to the drift equations are proportional to
the magnetic moment, thus deviations are expected to be minimal for
deeply-passing particles. Furthermore, the region of compressed
flux-surface is geometrically very small such that few NBI particles
are deposited in it. A low fraction of guiding-centres cross through
this area, similarly to field-lines avoiding the region (see figure
\ref{fig:heli_fieldlines}). Previous results on NBI fast particle
redistribution in the helical core of MAST is thus supported by this
full-orbit analysis. 

It is however reported that the investigation of fast particles with
larger Larmor radii (ICRH or fusion alphas) in similar MAST magnetic
configurations will probably require the use of higher-order GCDE or a
full-orbit approach. The switching techniques described in section
\ref{sec:switching} in conjunction with the field variation criteria
of section \ref{sec:field_variation} would greatly help completing
fast ion confinement simulations given moderate computational
resources.


\section{Conclusion}
\label{sec:conclusion}
In order to evaluate the limits of first order guiding-centre
approximation, a method for estimating the total perpendicular
variation of the magnetic field is proposed. It entails finding the
largest eigenvalue of a matrix, composed of covariant derivatives of
the magnetic field, projections and coordinate
transformations. Although lengthy to expand on paper, this operation
is easily performed on a computer and can be used within orbit
simulations to quickly asses the validity of first order GCDE. It is
also an instructive tool to probe the magnetic configuration for
understanding its characteristics, i.e. regions of strong gradients,
high curvature or large parallel currents.

A procedure to pass from guiding-centre to particle phase-space and
vice-versa is described, that is employed when a particle enters or
leaves a region of the magnetic configuration where the field
variation is above a given threshold. Numerical resources are thus
saved on particles with low energy for particles with larger Larmor
radii. The switching technique is suitable for slowing-down
simulations over long time-scales since the Monte-Carlo collision
operators dilute the small error caused by the switching. With this
algorithm, future simulations will yield more exact saturated fast ion
populations, optimally using computational resources.

The field variation criterion is applied to a case of heavily
distorted helical core equilibrium in MAST. It is reported that, for
energetic particles of around $10$ KeV, the proposed field variation
criterion reaches $12\%$ on the low-field side of the core and spikes
in the narrow region where the flux-surfaces are compressed. The
appearance of parallel currents explains the upsurge of field
variation, as the pitch of field-lines varies extensively within a
flux-surface and creates local shear.

Helical drift surfaces are obtained solving the particles' full-orbits
and tracing their guiding-centre trajectories. Confinement is
guarantied by the existence of drift surfaces inside the core and by
axisymmetry in the outer layers of the plasma. The kinked structure
impacts on the deposition of NBI fast ions, as co-passing particles
gather in the uncompressed flux-surface region. Full-orbit simulations
confirm that the guiding-centre formalism was sufficient to model the
NBI redistribution phenomena, mainly because NBI populations have
small Larmor radii and close to zero magnetic moment. The
investigation of more energetic and less tangential populations in
MAST require however the use of higher-order GCDE or a full-orbit
approach.

In future work, slowing-down simulations will benefit from the
switching technique to yield saturated fast ion distributions that are
exact even in strongly varying magnetic fields. The implementation of
second order corrections to the GCDE will result in a lighter system
of equations valid for intermediate field variations and an adjustment
to the proposed criterion will be sought.
%

%
\section*{Acknowledgements}
The authors are grateful to S. P. Hirshman for the use of the
\code{ANIMEC}/\code{VMEC} code for the equilibrium simulations, to
S. Brunner, L. Villard, D. Brunetti, J. Dominski and J. Faustin for
insightful discussions.
This work was supported in part by the Swiss National Science
Foundation.
%
\section{Appendix}
\label{sec:appendix}
\subsection{Charged particle orbits in a constant but sheared magnetic
  field}
\label{sec:orbit_sheared_field}
As an example of magnetic deformations that are completely ignored by
the leading order terms of the guiding-centre approximation, it is
instructive to solve the motion of charged particles in the constant,
curvature-free but sheared magnetic field $\bm{B}(\bm{x}) =
B_0\left[\sin(kx) \bm{e}_y + \cos(kx)\bm{e}_z\right]$. Up to a gauge
choice, the vector potential is $\bm{A} = \bm{B}/k$. Hence the
Lagrangian of such particle is written
\begin{multline*}
  \mathcal{L}(x,y,z,\dot{x},\dot{y},\dot{z})=\frac{1}{2}m v^2 + q \bm{A}\cdot\bm{v} \\
  =\tfrac{1}{2}m\left(\dot{x}^2+\dot{y}^2+\dot{z}^2\right) +
  \frac{qB_0}{k} \left[\dot{y} \sin(kx) + \dot{z}\cos(kx)\right]
\end{multline*}
The problem is integrable by virtue of the three constants of motion
that can be deduced from the symmetries of the above Lagrangian:
\begin{align*}
  \frac{\partial\mathcal{L}}{\partial y} &= 0 &\iff&& \frac{\partial\mathcal{L}}{\partial\dot{y}} = P_y &= m\dot{y} + \frac{qB_0}{k}\sin(kx) = const\\
  \frac{\partial\mathcal{L}}{\partial z} &= 0 &\iff&& \frac{\partial\mathcal{L}}{\partial\dot{z}} = P_z &= m\dot{z} + \frac{qB_0}{k}\cos(kx) = const\\
  \frac{\partial\mathcal{L}}{\partial t} &= 0 &\iff&& \mathcal{H} &=   \frac{1}{2}m\left(\dot{x}^2+\dot{y}^2+\dot{z}^2\right) = const
\end{align*}
Embracing the freedom of fixing the origin and orientation of the $y$
and $z$ axis, there is no loss in generality to consider the case
where, at $t=0$,
\begin{align*}
  x(0) &= 0  & y(0)&=\frac{mu_0}{qB_0}=\rho_0  & z(0) &= 0  \\
  \dot{x}(0) &= u_0  & \dot{y}(0) &= 0 & \dot{z}(0) &= v_0 
\end{align*}
i.e. $P_y=0$, $P_z=mv_0 + \frac{qB_0}{k}$ and
$\mathcal{H}=\frac{1}{2}m v^2= \frac{1}{2}m(u_0^2 +
v_0^2)$. The non-linear coupled system of equations describing the
motion of a charged particle respecting those initial conditions is
then
\begin{align}
  &\dot{y} = - \frac{qB_0}{mk} \sin(kx) = -\omega_0 \frac{\sin(kx)}{k} \label{eqn:sh_eom_y}\\
  &\dot{z} = v_0 + \frac{qB_0}{mk}\left[1 - \cos(kx)\right] =  v_0 +
  \tfrac{k}{2}\omega_0\frac{\sin^2(\tfrac{k}{2}x)}{\left(\tfrac{k}{2}\right)^2}\label{eqn:sh_eom_z}\\
  &\dot{x}^2 + \omega_0^2\left(1 + \frac{kv_0}{\omega_0}\right)\frac{\sin^2(\tfrac{k}{2}x)}{\left(\tfrac{k}{2}\right)^2} = u_0^2 \label{eqn:sh_eom_x}
\end{align}
where $\omega_0 = qB_0/m$ is the usual gyro-frequency. Equation
(\ref{eqn:sh_eom_x}) is similar to that of a pendulum, for which the
solution is entirely expressed in terms of elliptic integrals. For
$k\rightarrow 0$, equations (\ref{eqn:sh_eom_y}-\ref{eqn:sh_eom_x})
properly reduce to those of a charged particle in a constant magnetic
field pointing in the $z$-direction, for which the solution is the
well-known helical motion $x(t) = \rho_0\sin(\omega_0 t)$, $y(t)
=\rho_0\cos(\omega_0 t)$ and $z(t) = v_0 t$. In the special case
where $v_0 = -\omega_0/k$, the velocity along $x$ becomes constant,
i.e. $x(t)=u_0 t$ and the trajectory is circular in the $y-z$
plane, i.e. $y(t) = \tfrac{\omega_0}{k^2u_0} \cos(ku_0 t)$ and
$z(t)=-\tfrac{\omega_0}{k^2u_0}\sin(ku_0 t)$ with a radius of
$4\pi^2L^2/\rho_0$.

The motion in the $x$-direction is bounded ($\exists t_M \;|\;
\dot{x}(t_M)=0$) only if
\begin{gather}
  \left|\sin\left(\tfrac{k}{2}x_M\right)\right|=u_M =\frac{u_0}{\omega_0} \frac{k/2}{\sqrt{1+\frac{kv_0}{\omega_0}}} < 1\nonumber\\
  \iff \rho_0 <
  \frac{L}{\pi}\sqrt{1+\frac{kv_0}{\omega_0}},\label{eqn:shear_threshold}
\end{gather}
where $x_M$ is the maximum amplitude in the $x$-direction. This
condition demonstrates that if the Larmor radius $\rho_0$ is somewhat
larger than the characteristic length $L/\pi=2/k$ of field-line
shearing, particles do not necessarily perform a closed
gyro-motion. If they do, the period of the closed motion is
\begin{align*}
  T &= \int_0^T dt = \oint \frac{dx}{\dot{x}} 
  =\frac{4}{u_0}\int_0^{x_M}\frac{dx}{\sqrt{1 - \sin^2\left(\tfrac{k}{2}x\right)/\sin^2\left(\tfrac{k}{2}x_M\right)}}\\
  &= \frac{8u_M}{ku_0} \int_0^1 \frac{d\tau}{\sqrt{1-u_M^2\tau^2}\sqrt{1-\tau^2}}
  = \frac{4 K(u_M)}{\omega_0\sqrt{1+\tfrac{kv_0}{\omega_0}}} \overset{k\rightarrow 0}{\longrightarrow} \frac{2\pi}{\omega_0}
\end{align*}
where $K(k)$ is the complete elliptic integral of first kind. The
average velocity along the $Oz$ axis, which is the direction of the
magnetic field at the guiding-centre position, is
\begin{align*}
  <\dot{z}>&= \frac{1}{T}\int_0^T \dot{z}dt 
  =  v_0 + \frac{2\omega_0}{kT}\oint\frac{\sin^2\left(\tfrac{k}{2}x\right)}{\dot{x}} dx \\
  &= v_0 + \frac{16\omega_0 u_M^3}{k^2u_0 T} \int_0^1 \frac{\tau^2 d\tau}{\sqrt{1-u_M^2\tau^2}\sqrt{1-\tau^2}} \\
  &= v_0 + \frac{k u_0^2}{4\omega_0}\frac{1}{1+\tfrac{kv_0}{\omega_0}} \left[1 - \frac{E(u_M)}{K(u_M)}\right]\frac{2}{u_M^2}
\end{align*}
where $E(k)$ is the complete elliptic integral of second kind. By
expanding at leading order in the shearing parameter $k$, the correction
to the average particle motion along the parallel direction is
\begin{equation}
\label{eqn:z_correction}
<\dot{z}>-v_0 \overset{k\rightarrow 0}{\longrightarrow} \frac{ku_0^2}{4\omega_0}.
\end{equation}
The average parallel velocity is
\begin{equation*}
  <v_{||}>\ =\ < \bm{b}\cdot \dot{\bm{x}}>\ =\ < \sin(kx)\dot{y} + \cos(kx)\dot{z}>.
\end{equation*}
Using (\ref{eqn:sh_eom_y}-\ref{eqn:sh_eom_z}) and after some algebra,
the correction to the average parallel velocity is written
\begin{align}
<v_{||}> -v_0 
& = -\left(1+\frac{kv_0}{\omega_0}\right)\frac{2\omega_0}{kT}\int_0^T\sin^2\left(\tfrac{k}{2}x\right)dt\nonumber\\
&= -\frac{k u_0^2}{4\omega_0}\left[1 - \frac{E(u_M)}{K(u_M)}\right]\frac{2}{u_M^2} \overset{k\rightarrow 0}{\longrightarrow} -\frac{ku_0^2}{4\omega_0}.\label{eqn:para_correction}
\end{align}
Therefore, at leading order, there is a difference between the average
particle motion and the average parallel velocity
\begin{align}
<\dot{z}> - <v_{||}>= \frac{ku_0^2}{2\omega_0} = \frac{k \mu }{q} = \frac{\mu B}{m\omega_0}\bm{b}\cdot\left(\curl\bm{b}\right).
\label{eqn:banos_correction}
\end{align}
This paradoxical result is consistent with the so-called Baños drift
\cite[equation (37)]{banos}, for which an interpretation is given in
\citep[appendix B]{northrop-rome} as well as in
\citep[p.718]{cary-brizard}. In this example of magnetic fields with
no gradient nor curvature, the guiding-centre equations must be
extended with second order terms in order to encompass the shearing of
field-lines. In a Hamiltonian formalism, this can be performed
starting with the second order guiding-centre Lagrangian derived in
\cite[equations (32-33)]{littlejohn-1983},
\begin{multline*}
  \mathcal{L} = \left(q\bm{A} + mv_{||} \bm{b}\right)\cdot\bm{\dot{X}}\\
  - \left[\frac{1}{2}mv_{||}^2 + \mu B + \frac{1}{2}\frac{m}{q}\mu v_{||}
    \bm{b}\cdot(\curl\bm{b})\right],
\end{multline*}
which yields Euler-Lagrange equation that coincides with
(\ref{eqn:z_correction})
\begin{equation}
  \label{eqn:2nd_lagrangian}
  \bm{b}\cdot\bm{\dot{X}}= v_{||}  + \frac{1}{2}\frac{\mu}{q}\bm{b}\cdot\left(\curl\bm{b}\right).
\end{equation}
The guiding-centre motion along the field-lines is thus different from
the parallel velocity. The latter variable $v_{||}$ is interpreted in
the guiding-centre phase-space as a pivot variable that includes the
parallel curl of the magnetic field and the shearing of field-lines.


\subsection{Field variation estimation in various coordinates}
\subsubsection{Toroidal coordinates with geometric toroidal angle}
\label{sec:fieldvariation_curv}
The field variation can be estimated with algebraic expressions of
reasonable length in curvilinear coordinates where the toroidal angle
is equal to the geometric angle. Derivatives of the magnetic field can
be evaluated either in the covariant or the contravariant
representation. For example, the matrix of interest $\hat{M} = V^T G
V$ is written
\begin{align}
\label{eqn:Mdd}
M_{mn} &= V_{im}g^{ij} V_{jn} &  V_{im} &= B_{i;j}P^{jk}\frac{\partial x_m}{\partial u^k}
\end{align}
where $x_m = x^m = (x,y,z)$ are Cartesian coordinates, $u^k$
curvilinear coordinates. The perpendicular projection is the following
contravariant tensor
\begin{equation*}
  P^{jk} = g^{jk} - \frac{B^jB^k}{B^2},
\end{equation*}
the covariant derivative is
\begin{equation*}
  B_{i;j} = \partial_j B_i - \Gamma^k_{ij} B_k = \partial_j B_i - \Gamma_{ij,l} B^l,
\end{equation*}
and the Christoffel symbol of first type is used 
\begin{align*}
  \Gamma_{ij,l} = &\frac{\partial R}{\partial u^l}\frac{\partial^2 R}{\partial u^i\partial u^j} 
  + \frac{\partial Z}{\partial u^l}\frac{\partial^2 Z}{\partial u^i\partial u^j} \\
  &+ R\left[ \frac{\partial R}{\partial u^i} \delta_j^3\delta_l^3 + \frac{\partial R}{\partial u^j} \delta_i^3\delta_l^3 - \frac{\partial R}{\partial u^l}\delta_i^3\delta_j^3  \right].
\end{align*}
An other option is to lean on
\begin{align}
\label{eqn:Mud}
M_{mn} &= V^i_mg_{ij}V^j_n  &V^i_m &= B^i_{;j}P^{jk}\frac{\partial x_m}{\partial u^k}
\end{align}
where the covariant derivative is the mixed tensor
\begin{equation*}
  B^i_{;j} = \partial_j B^i + \Gamma^i_{jk} B^k
\end{equation*}
and the Christoffel symbol of second type is used
\begin{align*}
  \Gamma^i_{jk} = &\frac{\partial u^i}{\partial R}\frac{\partial^2 R}{\partial u^j\partial u^k} 
  + \frac{\partial u^i}{\partial Z}\frac{\partial^2 Z}{\partial u^j\partial u^k} \\
  & +\frac{1}{R} \delta^i_3\delta_j^3\frac{\partial R}{\partial u^k} + \frac{1}{R}\delta^i_3\delta_k^3 \frac{\partial R}{\partial u^j} - R\delta_j^3\delta_k^3\frac{\partial u^i}{\partial R}.
\end{align*}

In theory, the two options (\ref{eqn:Mdd}-\ref{eqn:Mud}) yield
identical results but in practice, computing equation (\ref{eqn:Mud})
is numerically more stable around the singular magnetic axis. In flux
coordinates, divergent metric terms such as $\partial^2 R/\partial
u^j\partial u^k$ are naturally multiplied by terms going to zero like
$\partial u^i/\partial R$ in the Christoffel symbol of second type
$\Gamma^i_{jk}$. Such cancellation happens later in the matrix
multiplication of (\ref{eqn:Mdd}) which makes it more prone to
truncation errors.
\subsubsection{Cylindrical coordinates}
\label{sec:fieldvariation_cyl}
In cylindrical coordinates $(r^1,r^2,r^3)=(R,\phi,Z)$, the metric is
simpler
\begin{align*}
  &\left|
    \begin{array}{l}
      x = R\cos\phi\\
      y = R\sin\phi\\
      z = Z
    \end{array}
  \right.  & \frac{\partial x_m}{\partial r^k} &=
  \begin{pmatrix}
    \cos\phi & -R\sin\phi & 0 \\
    \sin\phi & R\cos\phi & 0 \\
    0 & 0 & 1
  \end{pmatrix}
  \\
  g_{ij}&=
  \begin{pmatrix}
    1 & 0 & 0 \\
    0 & R^2 & 0 \\
    0 & 0 & 1
  \end{pmatrix}
  & g^{ij} &=
  \begin{pmatrix}
    1 & 0 & 0 \\
    0 & 1/R^2 & 0 \\
    0 & 0 & 1
  \end{pmatrix}.\nonumber
\end{align*}
The Christoffel symbols of first type $\Gamma_{ij,k} = \frac{\partial
  \bm{x}}{\partial r^k}\cdot\frac{\partial^2\bm{x}}{\partial
  r^i\partial r^j}$ are mostly zero except for
\begin{align*}
  \Gamma_{\phi\phi,R} & = -R & \Gamma_{R\phi,\phi} = \Gamma_{\phi R,\phi}=R \\
  \implies \quad \Gamma^R_{\phi\phi} &= -R & \Gamma^\phi_{R\phi} =
  \Gamma^\phi_{\phi R} = \frac{1}{R}.\nonumber
\end{align*}
Therefore the covariant derivative is written as
\begin{align*}
B^i_{;j} =
  \begin{pmatrix}
    \partial_R B^R & \partial_\phi B^R &  \partial_Z B^R \\
    \partial_R B^\phi & \partial_\phi B^\phi & \partial_Z B^\phi \\
    \partial_R B^Z & \partial_\phi B^Z & \partial_Z B^Z
  \end{pmatrix}
  +
  \begin{pmatrix}
    0 & -RB^\phi & 0 \\
    \frac{B^\phi}{R} & \frac{B^R}{R} &0 \\
    0 & 0 & 0
  \end{pmatrix}
\end{align*}
or alternatively
\begin{align*}
B_{i;j} =
  \begin{pmatrix}
    \partial_R B_R & \partial_\phi B_R &  \partial_Z B_R \\
    \partial_R B_\phi & \partial_\phi B_\phi & \partial_Z B_\phi \\
    \partial_R B_Z & \partial_\phi B_Z & \partial_Z B_Z
  \end{pmatrix}
  +
  \begin{pmatrix}
    0 & -RB^\phi & 0 \\
    -R B^\phi & R B^R &0 \\
    0 & 0 & 0
  \end{pmatrix}.
\end{align*}
The perpendicular projection matrix is expressed as
\begin{align*}
  P^{jk} &=
  \begin{pmatrix}
    1 & 0 & 0 \\
    0 & \frac{1}{R^2} & 0 \\
    0 & 0 & 1
  \end{pmatrix}
  -\frac{1}{B^2}
  \begin{pmatrix}
    {B^R}^2 & B^RB^\phi & B^RB^Z \\
    B^RB^\phi & {B^\phi}^2 & B^\phi B^Z \\
    B^RB^Z & B^\phi B^Z & {B^Z}^2
  \end{pmatrix}.
\end{align*}
%
%

\subsection{Small vector displacements in curvilinear coordinates}
\label{sec:curvcoord}
In Cartesian coordinates, $\bm{x} = \bm{X} + \bm{\rho}_L$ is simply
written $x^i = X^i + \hat{\rho}_L^i$, where $\hat{\rho}_L^i$ are the
Cartesian components of the Larmor radius vector. In curvilinear
coordinates, the coordinates are formally written $u^i(\bm{x}) =
u^i(\bm{X}+\bm{\rho}_L)$. This non-linear equation is simple to solve
if the mapping from Cartesian to curvilinear coordinates is
known. This mapping is not often provided and only the transformation
from curvilinear to Cartesian is available. Then, one relies on root
finding algorithms (slow) or linearisation (fast). For current
applications, it is sufficient to expand at lowest order in
gyro-radius, i.e.
\begin{align}
  u^i(\bm{x})\approx u^i(\bm{X}) + \bm{\rho}_L(\bm{X})\cdot\grad u^i
  = u^i(\bm{X}) + \rho_L^i(\bm{X})
  \label{eqn:curv_displacement}
\end{align}
where $\rho^j_L = \bm{\rho}_L\cdot\grad u^j$ are the components of the
Larmor radius in curvilinear coordinates. From prescription
(\ref{eqn:larmor_prescription}), those components are
\begin{align*}
\rho^i_L = \rho_L\frac{\bm{B}\times\grad
  B}{\left|\bm{B}\times\grad B\right|} \cdot\grad u^i
=\rho_L \frac{\epsilon^{ijk}}{\sqrt{g}}\frac{B_j}{B}\frac{\partial_k B}{N}
\end{align*}
where 
\begin{align*}
N&=\sqrt{\left|\grad B\right|^2 - \left(\bm{b}\cdot\grad B\right)^2} \\
&=\sqrt{(\partial_m B)g^{mn}(\partial_n B) - \left(B^l\partial_l B\right)^2/B^2}
\end{align*}

Flux coordinates are however singular near the magnetic axis and this
expansion can lead to wrong results when the norm of the basis vectors
diverge. For this reason, it is preferable to linearise in a
pseudo-Cartesian system. For example, if $u^1=\rho$ is a radial flux
label, $u^2 = u $ is a poloidal angle and $u^3 = v$ is a toroidal
angle, the pseudo-Cartesian coordinates are defined
\begin{align}
\left|
\begin{array}{l}
  \mathcal{X} = (R_0 + a \rho\cos u)\cos v\\
  \mathcal{Y} = (R_0 + a \rho\cos u)\sin v \\
  \mathcal{Z} = a \rho\sin u
\end{array}\right.
\label{eqn:pseudo-cart}
\end{align}
where $R_0$ is the major radius and $a$ the minor. Then, the small
displacement is written in the pseudo-Cartesian system as
\begin{align*}
  \mathcal{X}^i(\bm{x})\approx \mathcal{X}^i(\bm{X}) +
  \bm{\rho}_L\cdot\grad \mathcal{X}^i 
=\mathcal{X}^i(\bm{X}) + \rho_L^j(\bm{X})\frac{\partial
    \mathcal{X}^i}{\partial u^j}
\end{align*}
where the jacobian matrix $\frac{\partial \mathcal{X}^i}{\partial
  u^j}$ is easily obtained from (\ref{eqn:pseudo-cart}) and $\rho^j_L
= \bm{\rho}_L\cdot\grad u^j$ are again the components of the Larmor
radius in curvilinear coordinates (\ref{eqn:curv_displacement}). After
this operation, the $\mathcal{X}^i$ coordinates are easily inverted
back to the original curvilinear coordinates.


\bibliographystyle{phcpc}
\bibliography{biblio_ppcf}
\end{document}